\documentstyle[oldlfont,supertabular]{l-aa}
\input psfig

\newcommand{\nat}[2]{Nat #1, #2}
\newcommand{\apj}[2]{ApJ #1, #2}
\newcommand{\apjs}[2]{ApJS #1, #2}

\newcommand{\aj}[2]{AJ #1, #2}

\newcommand{\aaa}[2]{A\&A #1, #2}

\newcommand{\aeta}[2]{A\&A #1, #2}

\newcommand{\aetas}[2]{A\&AS #1, #2}
\newcommand{\pasp}[2]{PASP #1, #2}
\newcommand{\mn}[2]{MNRAS #1, #2}

\newcommand{\nh}{N$_{\rm H}$}
\newcommand{\ion}[2]{\hbox{#1\,{\small #2\normalsize}}}

\newcommand{\Halp}{H${\alpha}$}

\newcommand{\ergcm}{erg cm$^{-2}$ s$^{-1}$}
\newcommand{\ergs}{erg s$^{-1}$}

\newcommand{\degre}{\degr}
\newcommand{\degree}{\degr}
\newcommand{\Lbol}{\rm L$_{\rm bol}$}
\newcommand{\Lx}{$\rm L_{\rm X}$}
\newcommand{\rqvd}{r$_{90}$}

\newcommand{\cnts}{cts\ s$^{-1}$}
\newcommand{\lnls}{log\,N($>$S)-log\,S}
\newcommand{\tbsp}{\rule{0pt}{10pt}}
\begin{document}
   \thesaurus{06         
              (13.25.3;  
               13.25.5;  
               08.01.2;  
               08.14.1;  
               08.19.1;} 
   \title{The ROSAT Galactic Plane Survey: analysis of a low latitude sample area in 
Cygnus\thanks{Partly based on observations obtained at the Observatoire de Haute-Provence,
CNRS, France} 
    }
 
   \author{C. Motch \inst{1,2}
       \and    P. Guillout \inst{1}
       \and    F. Haberl   \inst{2}
       \and    M. Pakull   \inst{1}
       \and    W. Pietsch  \inst{2}
       \and    K. Reinsch  \inst{3}        
          }

   \institute{
              Observatoire Astronomique, UA 1280 CNRS, 11 rue de l'Universit\'e, 
              F-67000 Strasbourg, France
              \and 
              Max-Planck-Institut f\"ur Extraterrestrische Physik, D-85748,
              Garching bei M\"unchen, Germany 
              \and 
              Sternwarte, Geismarlandstrasse 11, D-37083, G\"ottingen, Germany
              }

   \date{Received 20 March 1996 / Accepted June 1996}

   \offprints{C. Motch}

   \maketitle

   \begin{abstract}

We present the analysis of the point source content of a low galactic latitude region
selected from the ROSAT all-sky survey. The test field is centered at $l$ = 90\degree , $b$ =
0\degree \ and has an area of 64.5\,deg$^{2}$. A total of 128 soft X-ray sources are detected
above a maximum likelihood of 8.  Catalogue searches and optical follow-up observations show
that in this direction of the galactic plane, 85\% of the sources brighter than
0.03\,PSPC\,\cnts \ are identified with active coronae.  F-K type stars represent 67\% ($\pm$
13\%) of the stellar identifications and M type stars account for 19\% ($\pm$ 6\%).  A small
but significant number of X-ray sources are associated with A type stars on the basis of
positional coincidence.  These results together with those of similar optical campaigns
demonstrate that the soft X-ray population of the Milky Way is largely dominated by active
stars.  We show that the density and distribution in flux and spectral type of the active
coronae detected in X-rays are consistent with the picture drawn from current stellar
population models and age dependent X-ray luminosity functions. The modelling of this
population suggests that most of the stars detected by ROSAT in this direction are younger
than 1 Gyr. This opens the possibility to extract in a novel way large samples of young stars
from the ROSAT all-sky survey. The small number of unidentified sources at low X-ray flux put
rather strong constraints on the hypothetical X-ray emission from old neutron stars accreting
from the interstellar medium. Our observations clearly rule out models which assume no
dynamical heating for this population and a total number of N$_{\rm ns}$ = 10$^{9}$ neutron
stars in the Galaxy. If accretion on polar caps is the dominant mode then our upper limit may
imply N$_{\rm ns}$ $\approx$ 10$^{8}$.  Among the non coronal identifications are three white
dwarfs, a Seyfert 1 active nucleus, two early type stars and one cataclysmic variable.  We
also report the discovery of a Me + WD close binary system with P$_{\rm orb}\ \approx \ $ 12
h.\footnote{Tables 11 and 12 are also
available in electronic form at the CDS via anonymous ftp cdsarc.u-strasbg.fr}

   \end{abstract}

%

\section{Introduction}

Starting in 1990 July the ROSAT X-ray satellite has performed during six months the
first soft X-ray all-sky survey ever made with an imaging instrument. At this
occasion, the position sensitive proportional counter (PSPC) was moved in the focal
plane of the X-ray telescope (XRT).  This detector offers an energy range of
0.1-2.4\,keV and an energy resolution of about 45\% at 1\,keV.  In survey mode,
the satellite was scanning the sky in great circles perpendicular to the solar
direction thus covering the whole celestial sphere in six months time. The spin
period of the satellite was synchronized with its orbital motion.  Any part of the sky
was continuously seen during a maximum detector crossing time of 32\,s every 96\,min.
The time interval during which a given source is visible mainly depends upon its
ecliptic latitude and the cumulative exposure time varies between 300 sec and 40,000
sec for the equatorial and polar regions respectively.  The mean flux sensitivity is
$\approx$ 2 10$^{-13}$ \ergcm \ (0.1-2.4\,keV). Taking into account the varying
point spread function across the PSPC field of view, the final spatial resolution of
the survey is of the order of 1.5\arcmin \ with a capability to localize point sources
with a 1$\sigma$ accuracy close to 20\arcsec.  A review of the main characteristics
of the ROSAT satellite and instrumentation can be found in Tr\"umper (1983) and
Pfeffermann et al. (1986) while the ROSAT all-sky survey (RASS) is described in Voges
(1992).

As part of the RASS the plane of the Galaxy was also surveyed entirely.  The analysis
of the region of the sky located below absolute galactic latitude 20\degree , known as
the ROSAT Galactic Plane Survey (RGPS) is the scope of a distinct project (Motch et
al. 1991).  Over 14,000 XRT survey sources are found in the Milky Way, most of these
being unknown before the launch of ROSAT.  

As for studies at high galactic latitudes, the RGPS source catalogue will serve as basis
for statistical analysis of various kinds of X-ray emitters. The galactic plane is
suited for studies of active coronae, OB stars and cataclysmic variables for
instance. It will also be extremely useful as finder for follow-up observations of
particularly interesting objects in the forthcoming years.  The RGPS may also
specifically address several pending questions of high astrophysical interest such as
the origin of the hard X-ray galactic ridge emission, the nature of supersoft X-ray
sources and the evolutionary status of accreting binary sources for example.

Compared with previous X-ray surveys of the Galaxy, the RGPS offers a much larger area
and improved sensitivity and spatial resolution.  The Einstein observatory mapped only
$\approx$ 2.5\% of the Galactic Plane (Hertz \& Grindlay 1984, 1988) at a flux limit
similar to that of ROSAT while the previous Milky Way surveys performed with collimating
instruments (e.g., HEAO: Nugent et al. 1983, Wood et al. 1984, EXOSAT: Warwick et al.
1985) had a sensitivity a factor 10 to 100 lower than that of the RGPS.

Obviously, the identification of this large amount of X-ray sources is far beyond the
nowadays possibilities.  Accordingly, we used three different paths for exploring this
unprecedented amount of new sources. The first and easiest step was the cross
correlation of RGPS source positions with astronomical catalogues, mostly extracted from
the SIMBAD database. The historical development of astronomy has favoured galactic
studies and the number of Milky Way objects held in catalogues is rather large. With the
positional accuracy of the survey, up to 30\% of RGPS sources may be readily identified
with SIMBAD objects with only limited rate of false coincidence (see e.g. Motch et al. 
1991, Voges 1992).  The second step consisted in selecting potentially interesting
objects for optical identification at the telescope. Using the observational RASS X-ray
properties of identified classes of X-ray sources together with other available
information at a different wavelength such as the Guide Star Catalogue of the Hubble Space
Telescope (Lasker et al. 1990) we could extract small subsets of sources having a higher
probability to be identified with a given kind of X-ray emitter than in the overall
survey installment. This approach led to several interesting results such as the
discovery of a luminous supersoft source in the Galaxy (Motch, Hasinger \& Pietsch 1994)
and of a new class of X-ray soft intermediate polars (Haberl et al.  1994, Haberl \&
Motch 1995).  The third step was the selection of several areas located in particular
directions of the Galaxy. Systematic optical identification of RGPS sources in these
small sized regions allows to estimate at least statistically the X-ray content of the
Milky Way as seen by ROSAT.  

Several groups working in both hemispheres participate into the optical identification
project in collaboration with scientists at the Max-Planck-Institut f\"ur
Extraterrestrische Physik. We report here on the efforts carried out by one of the
optical groups to optically identify ROSAT survey sources in a field located in the
galactic plane. In this paper we present the details of the identification strategy
and discuss the statistical properties of the sample which are likely to be
representative of the overall properties of the galactic ROSAT X-ray sky. We also
report on a few particularly interesting objects discovered in this field.  The
observed X-ray population of active coronae is compared with that predicted by
stellar population models having resolution in age, folded with the most recent X-ray
luminosity functions and the applications of such studies are shortly reviewed. We
also put constraints on the contribution to the galactic plane X-ray emission of an
hypothetical population of old neutron stars accreting from the interstellar medium.
Observational details and source finding charts are presented in a companion paper
(Motch et al. 1996a).

\section{The Cygnus area}

\subsection{Selection from the ROSAT all-sky survey}

The main scientific goal justifying the selection of this particular area in Cygnus
was the study of the X-ray source population at very low galactic latitudes. At the
time the optical observations started (1991 May) the Standard Analysis Software System
(SASS; Voges et al. 1992) had analyzed a limited fraction of the all-sky survey. The
only low galactic latitude areas then available were comprised between ecliptic
longitude $\lambda$ = 343\degree \ and $\lambda$ = 353\degree . Two years later, a
larger field embracing the former one was interactively analyzed using the Extended
Scientific Analysis System (EXSAS) developed at MPE (Zimmermann et al. 1992).  

As a result of this two step X-ray reduction process the completeness of the optical
investigations carried out in these two regions is slightly different and we shall
define an 'inner' area corresponding to the part in which the first automatic analysis
took place and a 'full' area corresponding to the EXSAS analysis.  

\begin{figure*}
\psfig{figure=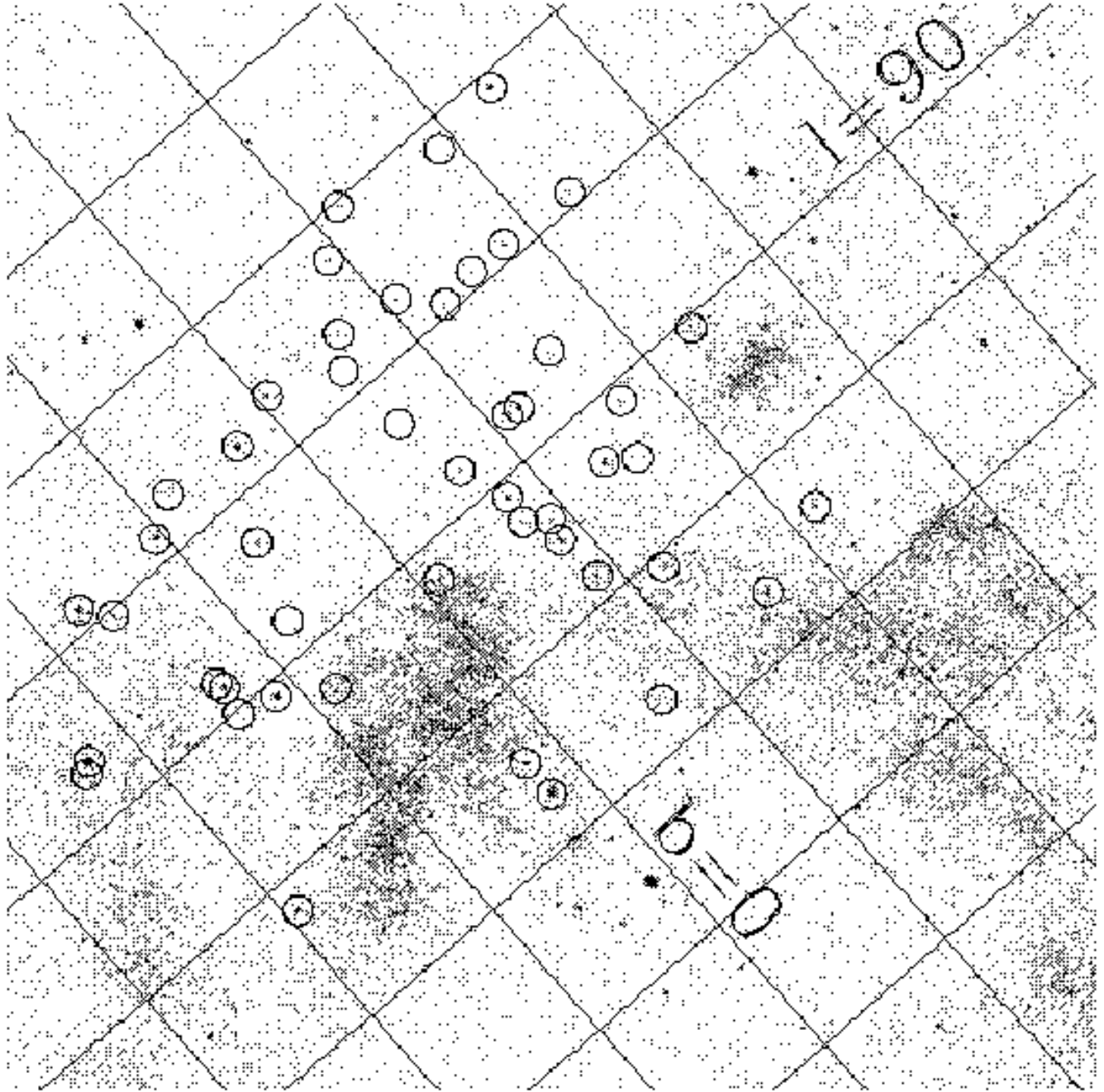,width=19cm,bbllx=3cm,bburx=18cm,bblly=6cm,bbury=21cm}

\caption[]{ROSAT all-sky survey image of the Cygnus sample region (0.5-2.0\,keV). The area is
centered on $l$ = 90\degree \ $b$ = 0\degree. The spacing of the coordinates grid is 2\degree
. Intense diffuse emission from the northern part of the Cygnus X-ray super bubble is clearly
seen. These notches of enhanced X-ray background were excluded from the final region which
totals 64.5\,deg$^{2}$.  Open circles mark the position of the point sources detected by the
EXSAS interactive analysis. Few very soft sources are not visible in the hard image.  For
clarity we only show here sources with count rate larger than 0.03\,\cnts}

\label{imageX}   
\end{figure*}

Another complication arises from the existence of intense diffuse emission from the
northern part of the Cygnus super bubble (see Fig. \ref{imageX}). The high number of
spurious point-like sources produced by both the SASS and EXSAS analysis in regions of
enhanced background and the resulting reduced sensitivity for detecting point sources
led us to discard these specific areas amounting to a total of 13.5\,deg$^{2}$.  

Both the 'inner' and 'full' areas are roughly centered at $l$ = 90\degree \ and $b$ =
0\degree .  The SASS analysis estimates the background from data collected within
4\degree \ wide strips parallel to the scan direction of the satellite in survey mode,
whereas the interactive analysis allows to handle in one run all X-ray photons
detected from the entire selected sky region.  This difference of approach means that
the source parameters, in particular, count rates, derived from the interactive run
are in principle more reliable than those given by the automatic process and we shall
not consider the latter in this paper.  Also, a slightly lower source acceptance
maximum likelihood (ML) for the 'full' area yielded the detection of several more
sources in the 'inner' area.  

\begin{figure}
\psfig{figure=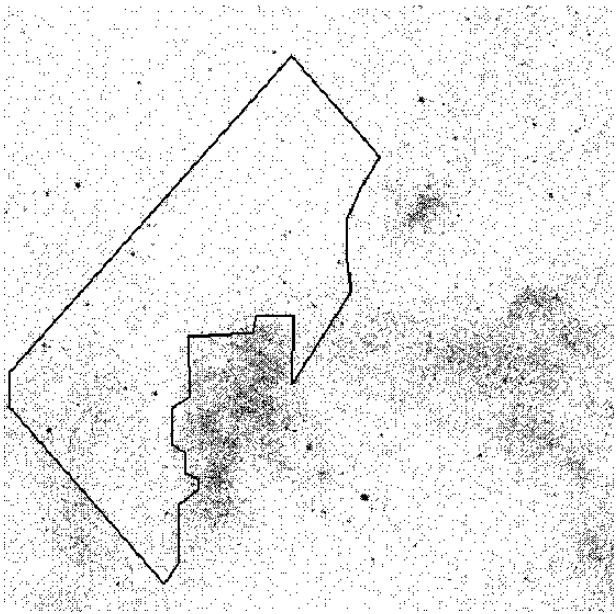,width=8.8cm,bbllx=3cm,bburx=18cm,bblly=6cm,bbury=21cm}
\caption[]{The 'inner' area over-plotted on the ROSAT all-sky survey image
 (0.5-2.0\,keV). The orientation is the same as in Fig.~1}
\label{sass_area}   
\end{figure}

\begin{figure}
\psfig{figure=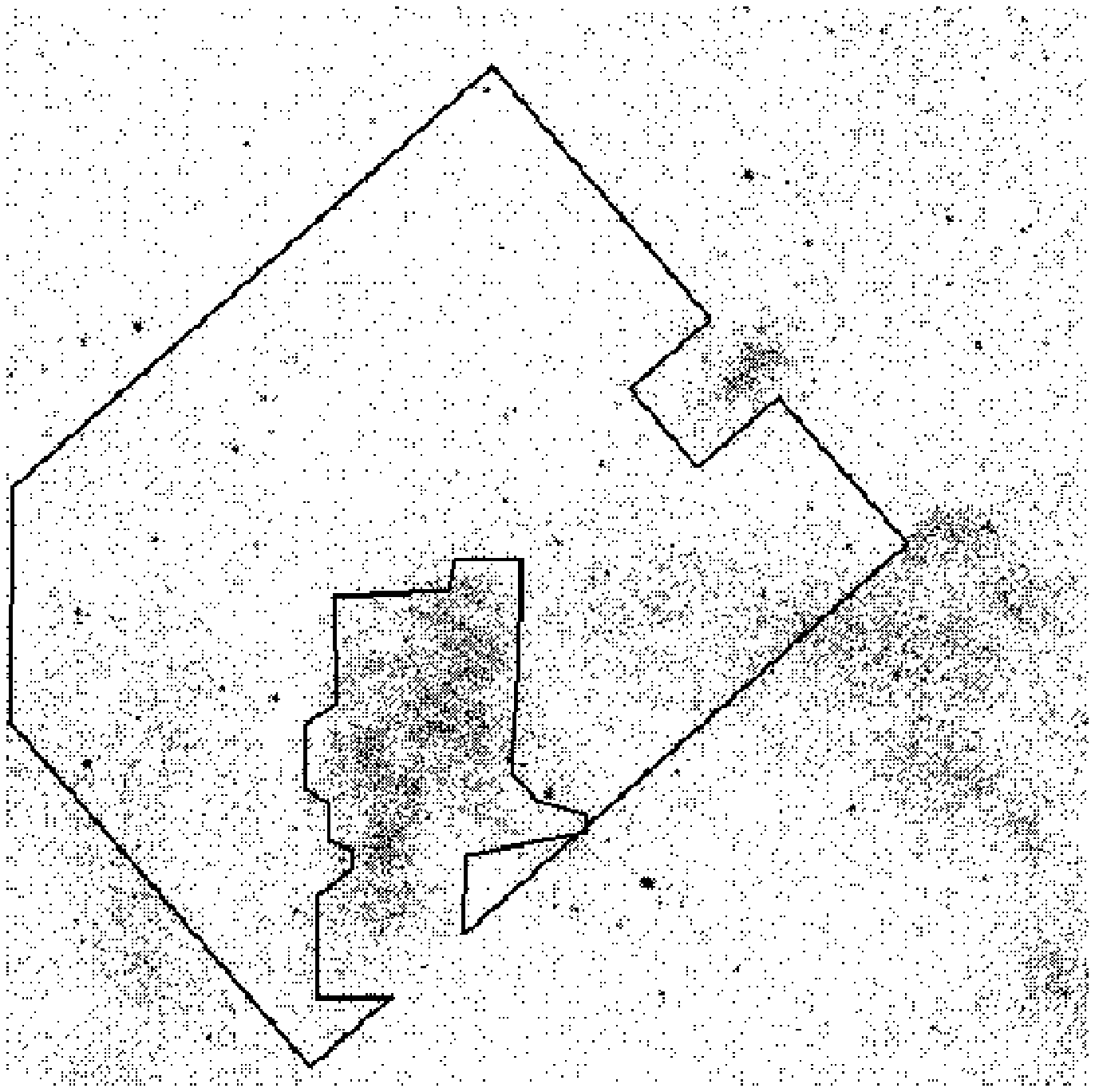,width=8.8cm,bbllx=3cm,bburx=18cm,bblly=6cm,bbury=21cm}
\caption[]{The 'full' area over-plotted on the ROSAT all-sky survey image
 (0.5-2.0\,keV). The orientation is the same as in Fig.~1}
\label{exsas_area}   
\end{figure}

The main features of the 'inner' and 'full' areas are listed in Table 
\ref{tab_areas}. We show on Figs \ref{sass_area} and \ref{exsas_area} the
positions of these two areas.

\begin{table}
\caption{Characteristics of the investigated areas}
\label{tab_areas}

\begin{tabular}{rrr}
                  &  'inner'                  & 'full'            \\
\hline
Area (deg$^{2}$)   & 38.1                   &  64.5               \\
\hline
Number of sources &                        &                     \\
ML $\geq$ 7       & 106                    & 158                 \\
ML $\geq$ 8       & 88                     & 128                 \\
ML $\geq$ 10      & 67                     & 95                  \\
\hline
\end{tabular}

\end{table}

The selected area is located in a range of galactic longitude which is closest to the
north ecliptic pole. Near the ecliptic poles the exposure time in the ROSAT all-sky
survey reached the maximum and therefore a strong gradient in exposure time is present
in the selected field from about 500\,s at $b = -5\degr$ to 1000\,s at $b = 5\degr$. 
However, our sample is rather homogeneous since 84\% of the sources have exposure
times ranging from 700 to 900\,s. Other areas in the galactic plane have lower
exposures, like e.g. the galactic center region with only a few hundred seconds (see
the exposure map of the ROSAT all-sky survey in Snowden et al.  1995).

The background level in the PSPC image, determined from source-free areas outside the
diffuse emission regions shows no strong variations and was between 0.7 and 0.8 counts
per square arcmin for the full energy band of 0.1 - 2 keV. The source detection using
the maximum likelihood technique from the EXSAS package was done in three energy
bands, namely 0.1 - 0.4 keV, 0.5 - 2.0 keV and 0.1 - 2.0 keV. Sources were formally
accepted above a maximum likelihood value of 7. Because of the high expected level of
spurious sources among the very low ML detections (up to 0.29 sources per square
degree or $\approx$ 19 spurious detections with ML $\geq$ 7 over the 'full' area, see
section 9.3) we only considered sources with ML $\geq$ 8 for most statistical
purposes.  Each of these sources, listed in Table \ref{optid}, has an associated
running index ranging from 1 to 128 and increasing with decreasing count rates. Since
some of the detections with ML between 7 and 8 had established optical counterparts of
scientific interest, we separately listed in Table \ref{optids} these additional
sources which were given index numbers in the range of 129 to 158.  

The area analyzed here is about 1/4 of that covered by the whole Einstein Galactic
Plane Survey (Hertz \& Grindlay 1984, 1988). However, the flux completeness level of
our sample in Cygnus is $\approx$ 4 times fainter and accordingly the total amount of
sources studied in this paper is comparable to that in the total Einstein survey.

\subsection{Astrophysical characteristics}

In the direction of $l$ = 90\degre , $b$ = 0\degree , the line of sight first crosses the
local spiral arm during the first 1\,kpc and then reaches the Perseus arm at a distance
of about 4\,kpc (e.g., Vogt \& Moffat 1975, Georgelin \& Georgelin 1976). 
Further away, the HI Cygnus arm (Kulkarni et al. 1982) is encountered at a distance of
$\approx$ 11\,kpc.

\begin{figure*}
\psfig{figure=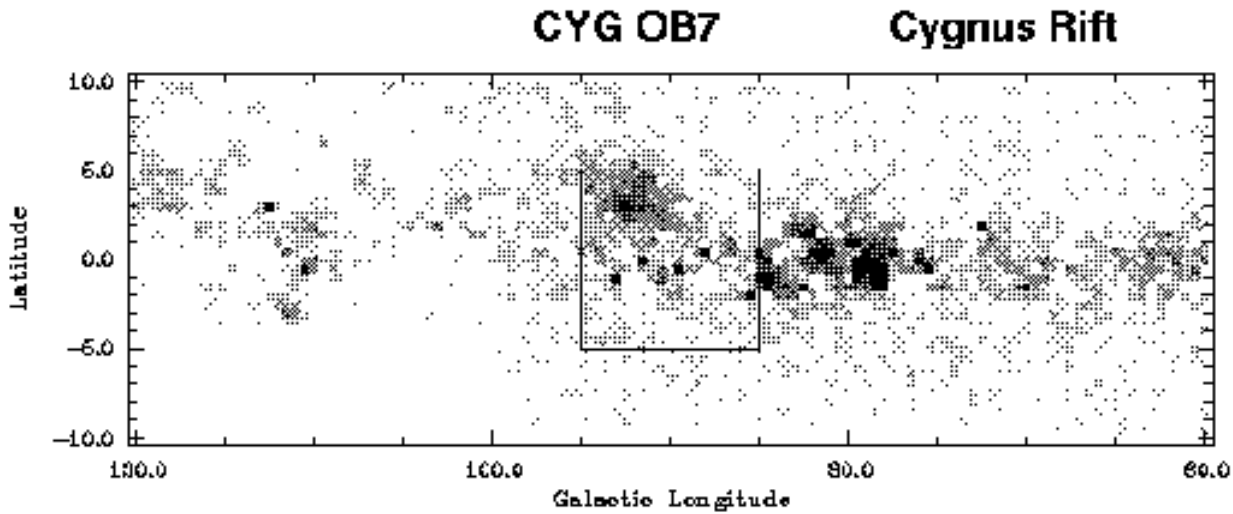,width=18cm,bbllx=0cm,bburx=14cm,bblly=0cm,bbury=8cm}

\caption[]{CO map (after Dame et al. 1987) of the galactic plane in a region surrounding
our ROSAT survey sample area. The square shows the approximate location of the area
studied in X-ray. In this region the dominant structure is the CYG OB7 molecular
cloud. A small part of the Cygnus Rift is also visible. These two complexes are located
at $\approx$ 700-800\,pc.  The peak of emission at $l$ = 80\degree \ is Cygnus X} 

\label{comap}   
\end{figure*} 

Two molecular cloud systems dominate the interstellar absorption and are well visible
on Fig. \ref{comap}. At $l$ $\leq$ 87\degree \ the edge of the Cygnus Rift cloud complex
appears.  However, the main structure is the cloud related to the CYG OB7 association
(Dame \& Thaddeus 1985) located in the $l$ $\geq$ 90\degre \ ; $b$ $\geq$ 0\degree\ 
part of our selected region.  From CO measurements Dame \& Thaddeus (1985) derive
distances of 700 - 800\,pc for these two molecular clouds, well within the local spiral
arm.  These two structures are also visible in the optical absorption map of Neckel \&
Klare (1980) at a comparable distance.  Finally, a similar structure may be seen in
the dark cloud map of Feitzinger \& St\"uwe (1986).  

Our sample region overlaps with the north-east part of the X-ray super-bubble discovered by
Cash et al. (1980). This ring-shaped soft X-ray diffuse emission is well seen in Snowden
et al. (1995). This hot bubble is thought to be located at a distance of about 2\,kpc and
its total angular extent of 13\degree \ implies a diameter of 450\,pc. The total X-ray
energy radiated is 5 10$^{36}$\,erg s$^{-1}$ at a temperature of 
2 10$^{6}$\,K and the total energy content of the bubble is estimated to be of the order of
10$^{51}$ ergs (Cash et al. 1980). The origin of this large structure is unknown but
probably related to the Cyg OB2 association. A series of 30-100 supernovae during the last
3-10 million years or hot winds emanating from OB associations and interacting with the
interstellar medium could explain the X-ray emission.

\section{Optical identification procedure}

\subsection{SIMBAD identifications}

Our first step toward optical identification was to search the SIMBAD database for
positional coincidence with a catalogued object. For the entire galactic plane survey,
about 20\% to 30\% of the sources may be readily identified with SIMBAD entries with
an expected number of random spurious coincidence of about 10\% of the
identifications (Motch 1992). A similar identification rate was derived from the
analysis of a small area in the galactic plane in Perseus (Motch et al. 1991).

In our 'full' area, 30 among 95 sources (i.e. 31\%) detected at a maximum likelihood
larger than 10 have SIMBAD entries within \rqvd \ the 90\% ROSAT survey confidence
radius. Scanning the database shows that the mean density of SIMBAD objects in a
10\degree  $\times$10\degree \ region centered on $l$ = 90\degree , $b$ = 0\degree \
is $\approx$ 130 per square degree. With an average \rqvd \ = 25\arcsec \ (ML $\geq$
10),  we expect a random match in $\approx$ 2\% of the cases. This lower false
identification rate compared to former studies is due to the improved accuracy of
survey X-ray sources positioning with respect to the first analysis.  We thus expect 2
wrong identifications among the 30 SIMBAD proposed matches. However, the spatial
density of SIMBAD entries may be locally much higher because of the inclusion of a
particular catalogue (e.g. the catalogue of objects in the direction of M39 compiled by
Platais, 1994).  Therefore, whenever the information retrieved from the literature
was not firmly conclusive we tried to obtain our own optical data.  

\subsection{Optical observations}

All optical data were obtained at the Observatoire de Haute-Provence in the time
interval from 1991 May till 1993 September. A description of the instrumentation used
may be found in Motch et al. (1996a).  

For each ROSAT source we produced a finding chart using the HST Guide Star Catalogue
(GSC, Lasker et al. 1990). GSC data were extracted in a first step from the STARCAT
facility at ESO (Pirenne et al. 1993) and later from the SIMBAD catalogue GSC browser
(Preite-Martinez \& Ochsenbein 1993).

In the cases where one or more rather bright GSC stars were lying close to the center
of the X-ray position, with apparently a rather low probability of random coincidence,
we directly started our investigation by obtaining medium resolution spectroscopy
($\lambda \lambda$ 3800 - 4300 \AA ; FWHM resolution $\approx$ 1.8 \AA ) of the GSC
candidates. These spectra allowed the detection of re-emission in the \ion{Ca}{II}\ H\&K lines
which is a well known signature of chromospheric and associated coronal activity (e.g.
Schrijver 1983, Maggio et al. 1987).  When the chromospheric activity was not found at
the level expected from the intensity of the X-ray source (see below) or when no
bright GSC star was conspicuous in the error circle, we obtained B and I band CCD
photometry of the field.  From the B-I index we could select stars exhibiting a red
excess, presumably dM star candidates and blue excess objects considered as candidates
for white dwarfs, cataclysmic variables and background AGNs. We then obtained low
resolution spectroscopy ($\lambda \lambda$ 3500 - 7500 \AA ; FWHM resolution $\approx$
14 \AA) of the photometric candidates and of other faint sources if necessary. We
tried to push our optical low resolution spectroscopic investigations at the same
limiting magnitude for all sources at a given X-ray count rate in order to preserve as
much as possible the completeness of the identified sample. This means that for the
X-ray brightest unidentified sources we are complete down to V $\approx$ 17-18.
However, considering the sometimes relatively high number of objects encountered in
the error circles and the changing instrumental conditions our optical depth is
probably not quite homogeneous.

\section{Optical identifications}

\subsection{Extragalactic sources}

Active galactic nuclei account for almost 70\% of all extragalactic identifications in
the Einstein Extended Medium Sensitivity Survey (EMSS; Stocke et al. 1991) which has a
median sensitivity close to that of the ROSAT all-sky survey. With this X-ray
sensitivity AGNs have reddening free V magnitudes in the range of 14 to 20 (Gioia et
al.  1984).  Redshifted broad and narrow emission lines are easily recognizable
optical signatures even at low signal to noise ratio. From HI and CO data we find that
the clearest part of the 'full' region has only $\rm A_{V}$ $\approx$ 2-3 mag whereas
the most absorbed part has $\rm A_{V}$ $\geq$ 15 mag. Therefore we do not expect to
identify many AGNs with our instrumental spectroscopic limit of V $\approx$ 18 and may
consider our only identified AGN at V = 16.3 as outstanding.  The situation is even
more unfavourable for clusters of galaxies which constitute the second most numerous
class of extragalactic sources.  Their intrinsic optical faintness and the high
galactic stellar foreground make their recognition on a CCD image almost impossible in
such a low latitude region. Finally the lack of marked optical signatures renders
identification of BL Lac objects probably hopeless.

\subsection{White dwarfs}

Because of the very pronounced effect of interstellar absorption on their intrinsically
very soft X-ray spectra, most of the white dwarfs detected by ROSAT
undergo very little interstellar absorption and thus appear basically unreddened at
optical wavelength.  The hottest DA and DO members which are the most likely to be
serendipitously discovered by ROSAT exhibit only weak and shallow H or He absorption
lines (Wesemael et al. 1993) which are sometimes difficult to detect for the optically
faintest objects. Therefore, we rather considered the very blue Rayleigh-Jeans like
continuum as the definite signature of a white dwarf.  Of the three white dwarfs
detected in our sample area one was already catalogued as such (GD 394).  The optically
brightest of the two new white dwarfs was found recently in a list of UV excess
objects in the galactic plane (LAN 121; Lanning \& Meakes 1994).

\subsection{Early type Stars}

Einstein X-ray observations have shown that early type stars may have soft X-ray
luminosities as large as a few 10$^{33}$ \ergs . Pallavicini et al. (1981)
demonstrated that the soft X-ray and bolometric luminosities of O-B stars were tightly
correlated with \Lx \ $\approx$ 10$^{-7}$\,\Lbol . Among the $\approx$ 300 stars
earlier than B5 listed in SIMBAD for our Cygnus survey area, we detect two of the
three optically brightest (B $\leq$ 6.0) (HR 8154, B = 4.99, O8e and HD200310, B =
5.16, B1Ve) but we miss the overall brightest one (HR 8047, B = 4.69, B1.5nne).  These
stars have photometric distances of the order of 600\,pc and are probably members of
the CYG OB7 association. In these two cases, the X-ray luminosities are
consistent with normal coronal emission from the OB star. Our third detection is HR
8106 which is a B9III/Ap star rather than a hot early type star and the inferred X-ray
luminosity (\Lx \ $\approx$ 1 10$^{30}$ \ergs) is consistent with that of a late type
active companion star.  

\subsection{Active coronae}

The fact that many late type stars were bright soft X-ray sources was one of the
important discoveries of the Einstein observatory (Vaiana et al. 1981). X-ray coronal
activity is now known to be a common feature of many stars of spectral type later than
about A5 (e.g. Rosner et al.  1985 and references therein).  Observed stellar X-ray
luminosities are in the range of 10$^{26}$-10$^{31}$ \ergs . They increase with
rotation rate (Pallavicini et al. 1981) and decrease with stellar age (Micela et al.
1988), pre main sequence stars being about 1000 times more luminous than old main
sequence objects.  Although a satisfactory theory accounting for all aspects of
stellar X-ray activity is not yet available, the currently accepted picture is that
X-ray emission originates from a hot stellar corona. By analogy with what is known
from the Sun the bulk of X-ray emission is thought to come from magnetically confined
loop-like structures emerging at the stellar surface and which are likely to be
generated in the subphotospheric convective layers by a dynamo mechanism.  

Unfortunately, there is no bright spectral signature of coronal activity at optical
wavelength. However, chromospheric and coronal activities are known to be well
correlated in the Sun and late type stars in general (see e.g. Reimers 1989 and
references therein).  In particular, the \ion{Ca}{II} H\&K chromospheric emission lines are
very sensitive measurements of stellar activity which behaves like X-ray emission in
being more intense in fast rotating and young stars (Wilson 1963, 1966).  Studies
based on Einstein X-ray data have shown that the strength of these emissions is indeed
well correlated with X-ray luminosity (e.g.  Schrijver 1983, Maggio et al. 1987,
Fleming et al. 1988).  

\subsubsection{Optical spectroscopy}

Using medium resolution blue optical spectra of $\approx$ 100 stars associated with RGPS
sources Guillout (1996) has established and calibrated the relation between \ion{Ca}{II} 
H\&K and PSPC flux. Basically, \ion{Ca}{II} H\&K and soft X-ray emissions are found to be
correlated both in flux and luminosity with F$_{\rm CaII} \ \sim $ F$_{\rm X}^{0.74 \,
\pm \, 0.14}$ and L$_{\rm CaII} \ \sim$ L$_{\rm X}^{1.05 \, \pm \, 0.20}$. These
relations hold over 2 decades in flux and over 3 decades in luminosity. The existence
of a flux/flux correlation leaves no doubt that the luminosity relation is real and not
an artifact of the distribution in distance. A rms scatter of about a factor 2 in flux
/ luminosity exists around the mean trend with a maximum range of a factor 10. This
scatter is probably caused by the long term (solar cycles) and short term (flares)
variability of the coronal/chromospheric activity since our spectroscopic measurements
were obtained on the average 2 years after the X-ray survey observations.  In spite of
this dispersion the above relations may still be extremely valuable tools for
identifying active coronae.  

\begin{figure}
\psfig{figure=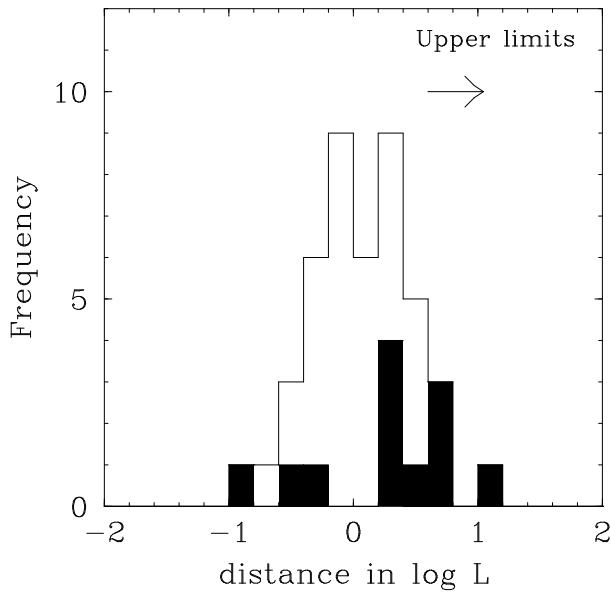,width=8.8cm,bbllx=1cm,bburx=19cm,bblly=1cm
,bbury=19cm}

\caption[]{Histogramme of the distance to the best log (L$_{\rm CaII}$) / log (L$_{\rm
X}$) relation for 44 candidate stars in Cygnus (measured values and upper limits). The
filled histogram represents stars for which only an upper limit on \ion{Ca}{II} H\&K emission is
available. Stars with positive log distances exhibit less \ion{Ca}{II} emission with respect to
their X-ray flux than for the average identified sample and on this basis may be
considered as less likely counterparts}

\label{deltaxca}   
\end{figure}

The distances in the log (L$_{\rm CaII}$) / log (L$_{\rm X}$) diagram between the
position of each active corona and the mean relation have an apparently Gaussian
distribution with $\sigma$ = 0.30 (Guillout 1996).  We show in Fig.
\ref{deltaxca} the distribution of these distances for our sample in Cygnus.  In order
to quantify the \ion{Ca}{II} identification criterion we computed for each star P$_{\rm CaII}$,
the formal probability that if the star is responsible for the X-ray emission its
chromospheric emission appears fainter or equal to the observed value.

\begin{figure}
\psfig{figure=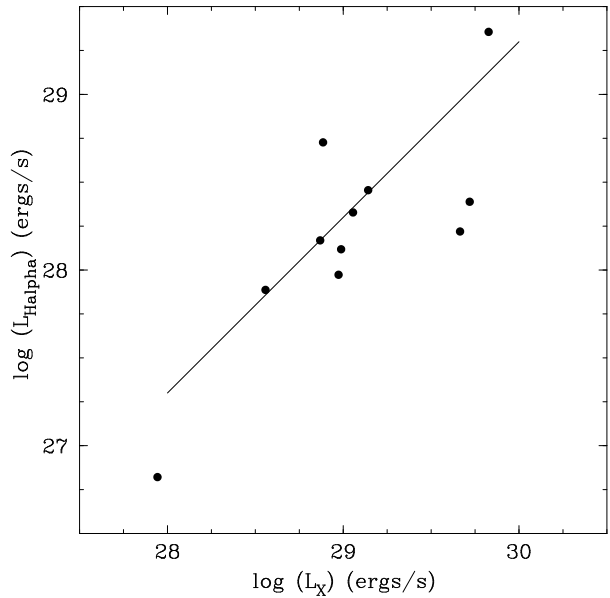,width=8.8cm,bbllx=2cm,bburx=19cm,bblly=2cm,bbury=19cm}

\caption[]{Comparison of chromospheric \Halp \ and coronal X-ray luminosities for
11 Me stars identified in the 'full' area and having a low resolution spectrum. The
solid line represents the relation found by Skumanich et al. (1984) which is
also representative of the EMSS Me stars} 
\label{mec}   
\end{figure}

Several M stars which were optically too faint for medium resolution spectroscopy were
observed at low resolution ($\lambda \lambda$ 3500- 7500 \AA ; FWHM resolution
$\approx$ 14 \AA ).  In these cases, we considered that the detection of Balmer
emission unambiguously identifies the X-ray source with the late type star. It is known
that active M dwarfs tend to have stronger Balmer over \ion{Ca}{II} H\&K emission ratios than
earlier types (Rutten et al. 1989) and that H$\alpha$ and X-ray emissions tightly
correlate (Fleming et al. 1988). In order to check the validity of our Me star
identifications we computed \Halp \ fluxes using our spectra and the V magnitudes
extracted from the literature or from our own measurements. X-ray luminosities were
estimated assuming F$_{\rm X}$ $\approx$ 10$^{-11}$ $\times$ S \ergcm \ where S is the
PSPC count rate in \cnts . For RX\,J2104.1+4912 (index 7) we used the quiescent X-ray count rate
(see sect. 6). In the \Lx \ / L$_{\rm H\alpha}$ diagram (Fig. \ref{mec}) our Me stars occupy
a region similar to that occupied by the Me stars identified in the EMSS (Fleming et
al.  1988) or by Skumanich et al. (1984). Considering the remaining photometric errors
of $\approx$ 0.3 mag and keeping in mind the slightly different X-ray energy ranges
between ROSAT and Einstein we consider that the agreement is good and leaves no doubt
that we have correctly identified these X-ray sources.  

\subsubsection{Positional coincidence with GSC}

\begin{figure}
\psfig{figure=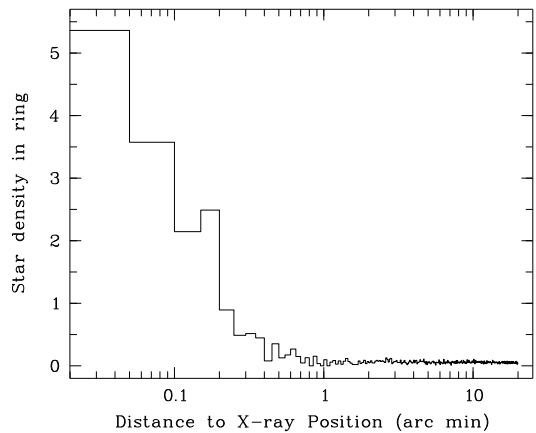,width=8.8cm,bbllx=1cm,bburx=27cm,bblly=-1cm
,bbury=20cm}

\caption[]{Mean density of GSC entries (arcmin$^{-2}$) in a ring of 3\arcsec \ width
centered on the X-ray position as a function of the radius of the ring. The density
shown here is the average for 95 sources of the 'full' area having a maximum
likelihood larger than 10.  The clear excess of matches at small distances
demonstrates the possibility to automatically identify sources with GSC entries (or
at least select active coronae candidates)}

\label{fig_gsc_rayleigh}   
\end{figure}

Visual examination of our Guide Star Catalogue finding charts readily showed that in many
cases the X-ray position was quite close to a rather bright (typically V $\leq$ 12)
star usually unreferenced in SIMBAD. In order to obtain an additional identification
criterion we computed the a priori probability of positional coincidence by
systematically scanning the GSC catalogue within 20\arcmin \ from the ROSAT position.
Care was taken to remove duplicate stars in the regions of plate overlaps. We show on
Fig.  \ref{fig_gsc_rayleigh} the density of GSC entries held in a ring-shaped area
centered on the X-ray position as a function of the radius. The high density excess of
stars close to the X-ray position shows that a sizeable fraction of our X-ray sources
have indeed counterparts in the GSC catalogue. For instance, 59 GSC entries are found
within 30\arcsec \ from the X-ray position whereas the expected number of random
matches is only 4. This example shows that in the galactic plane at least, systematic
cross-correlation of the ROSAT all-sky survey sources with GSC entries should allow
efficient source selection if not identification.

For each X-ray position with a GSC entry of magnitude $V$ within \rqvd \ we defined
the a priori probability of a spurious match within \rqvd \ as P$_{\rm spurious}$  =
N$_{\rm V} \times (\frac{r_{90}}{20 '})^{2}$ where N$_{\rm V}$  is the number of GSC
entries of magnitude brighter or equal to $V$ found within 20\arcmin \ from the ROSAT
source.  In a similar fashion we define the a priori probability of identification of
the X-ray source with the GSC star as P$_{\rm GSC} = 1\ - \ $P$_{\rm spurious}$. When
no GSC entry was found within \rqvd \ we set  P$_{\rm GSC}$ = 0.0 and used the
\ion{Ca}{II} H\&K emission alone as identification criterion. 

\begin{figure}
\psfig{figure=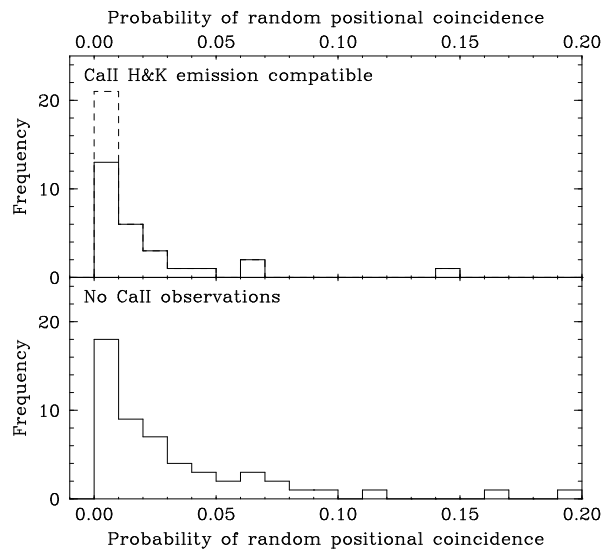,width=8.8cm,bbllx=1cm,bburx=21cm,bblly=2cm
,bbury=20cm}

\caption[]{Histogramme of the probabilities of random positional coincidence of X-ray sources
with GSC entries. Upper panel: all stars having a measured \ion{Ca}{II} H\&K emission compatible
with ROSAT X-ray flux, solid line; only detection, dashed line; including compatible upper
limits. Lower panel: proposed candidate stars without spectroscopic observations. The two
distributions are similar and the expected total number of false coincidence is 0.6 among
35 for the spectroscopic sample and 1.7 among 53 for the positional sample. This
demonstrates that in the galactic plane one may identify ROSAT survey sources with GSC
entries with a high success rate on the basis of positional coincidence only}

\label{de90}   
\end{figure}

We show in Fig. \ref{de90} the histograms of P$_{\rm spurious}$ for the samples of
candidate stars having a \ion{Ca}{II} measurement compatible with the X-ray emission (i.e. 
with a probability of more than 2\% to belong to the mean \ion{Ca}{II} / X luminosity
relation) and for those proposed on the only basis of their proximity to the X-ray
position.  The two distributions look alike and by integrating the probabilities one
expects in total 0.6 among 35 and 1.7 among 53 false cases in the spectroscopic and
positional selected samples respectively. This gives us confidence in the possibility
to identify X-ray sources with rather bright GSC entries, all presumably active coronae,
without obtaining time consuming medium resolution spectroscopy.

Because of the late interactive re-analysis of X-ray survey data in this field, we
could spend significantly more observing time per source in the 'inner' area than in
the 'full' area.  Consequently,  the majority of our spectroscopic sample is in the
'inner' area. For the sources located outside the 'inner' area, we concentrated on
X-ray locations without bright GSC candidates relying on the probability of positional
coincidence to identify the un-investigated X-ray sources.

Note that for convenience we arbitrarily assigned a position probability of 1.0 to our
optical identifications with AGN, white dwarfs and Me and late Ke stars. A couple of stars
with GSC positions inside \rqvd \ (sources index 13 and 80) were eventually found slightly
outside when using SIMBAD coordinates.

\subsubsection{Combined identification criteria}

The positional and \ion{Ca}{II} probabilities have different natures since we are dealing on
one hand with the probability that a bright unrelated star falls into the ROSAT error
circle and on the other hand with the probability that a given star is actually
related to the X-ray source once it is found close to the ROSAT position. As a rule of
thumb we decided to set the probability boundaries such as the number of spurious or
missed identifications among the considered samples was of the order of 1.  Since in
the 'full' area, we have 44 candidate active coronae with a GSC match and no
spectroscopy available and the same number of candidate stars with spectroscopic
observations, we decided to identify with active coronae all X-ray sources having
either P$_{\rm GSC}$ $\geq$ 98\% or P$_{\rm CaII}$ $\geq$ 2\%. Stars without measured
\ion{Ca}{II} flux or with P$_{\rm CaII}$ $<$ 2\% and 95\% $\leq$ P$_{\rm GSC}$ $<$ 98\% were
considered as potential optical counterparts and marked with a '?' in columns 'Class'
and 'Identification' of Table \ref{optid} and \ref{optids}. We count 13 such 95\% confidence level
identifications with ML $\geq$ 8 whereas only $\approx$ 4 such random associations are
expected for a set of 128 sources. This illustrates further the usefulness of the
Guide Star Catalogue for identifying RASS sources at low galactic latitudes.

\begin{figure}
\psfig{figure=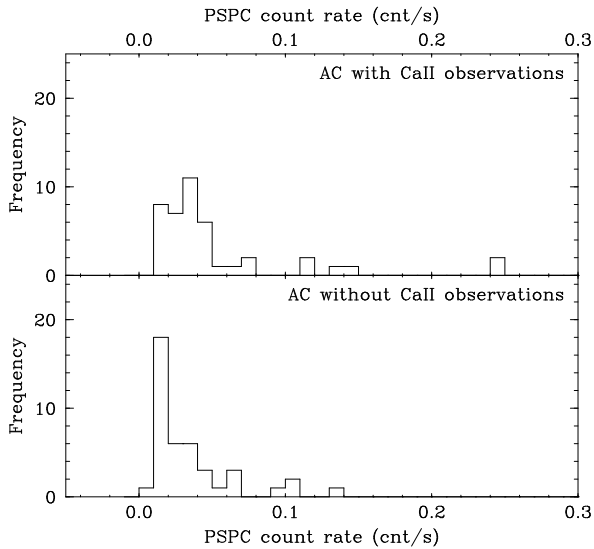,width=8.8cm,bbllx=1cm,bburx=21cm,bblly=2cm
,bbury=20cm}

\caption[]{Histogramme of the PSPC count rates for active coronae identified in the
'full' area on the basis of the strength of their \ion{Ca}{II} H\& K emission (upper panel)
and on the basis of positional coincidence only (lower panel). The difference of the
source distribution reflects our optical identification strategy}

\label{fxcaii}   
\end{figure}

We list in Table \ref{acstat1} the statistics of the origin of identifications with
active coronae for the original 'inner' and 'full' areas. The distribution in count
rates of the two identified samples is shown in Fig. \ref{fxcaii}.

\begin{table}

\caption{Origin of active coronae identifications for sources with maximum likelihood
larger than 8}

\label{acstat1}

\begin{tabular}{lrr} 
Criterion of identification                          &  'inner'     & 'full' \\ 
\hline 
\ion{Ca}{II} H\&K emission (P$_{\rm CaII} \ \geq$ 2\%)          &  32         & 40 \\ 
Low resolution spectroscopy                          &   8         &  9 \\
Positional coincidence only (P$_{\rm GSC} \ \geq $ 98\%) &  16         & 25\\ 
\hline 
Total number of identified  active coronae           &  56         & 74\\ 
\end{tabular} 
\end{table} 

Among the 44 candidate stars located in the 'full' area and having medium resolution
spectra, we find 4 cases (all upper limits) where the probability that the observed
\ion{Ca}{II} emission is compatible with X-ray emission is smaller than 2\% whereas the
normal distribution would predict only one such case.  For two of these sources
(RX\,J2055.3+5025, index 15 and RX\,J2054.1+4942, index 44) the optical counterpart is
bright (V $\leq$ 11) and the a priori chance probability of positional random coincidence
derived from the GSC surroundings is correspondingly very small, typically less than 1\%. 
These two bright candidates of spectral types F8 and G0 have strong photospheric
\ion{Ca}{II} flux and any emission will have lower contrast than in later type stars. 
Velocity differences between emission and absorption components produced by fast rotation
or binarity will further decrease the \ion{Ca}{II} contrast.  In the two remaining cases
(RX\,J2132.5+4849, index 51 and RX\,J2054.6+5120, index 72), the probability of random
coincidence is much larger ($\geq$ 5\%) and we shall not consider these two
identifications.  All 8 stars for which we only had \ion{Ca}{II} upper limits compatible
with the observed ROSAT X-ray flux also had P$_{\rm GSC}$ $\geq$ 98\% and on this basis
were considered as reliable optical identifications.

\section{Overall source statistics}

\subsection{Completeness of identification and repartition by types}

The final list of proposed optical counterparts including the positional and \ion{Ca}{II} derived
probabilities of identification are listed in Tables \ref{optid} and \ref{optids} .  We
show in Tables \ref{overallstat1} and \ref{overallstat2} the statistics of source
identifications in the 'full' area for several limiting count rates.  For sources detected
with ML $\geq$ 10, the X-ray completeness level is $\approx$ 0.02\,\cnts \ (see section 9)
and at this limit we have optically identified 88\% of the sources.  Active coronae
constitute the overwhelming majority of our sample. The contribution of white dwarfs
reduces the stellar fraction at high count rates and the incompleteness of the optical
identification again decreases the fraction of active coronae at low count rates. At the
level of 0.03--0.025\,\cnts \ which is about the median all-sky survey sensitivity we have
identified over 95\% of the sources and active coronae account for about 85\% of the total
population in our low galactic latitude field.  A high fraction of active coronae was also
found in the Einstein Galactic Plane Survey ($\approx$ 46\%; Hertz \& Grindlay 1984, 1988)
and in a ROSAT sample area in Perseus ($\approx$ 48\%; Motch et al. 1991).  Our record
beating percentage of active coronae in Cygnus obviously reflects the location at $b$ =
0\degree \ of our area. A similar test region in the Taurus Constellation, away from the
main star forming region ($b$ $\approx$ -14\degree ; $l$ $\approx$ 180\degree ) also
contains $\approx$ 77\% of active stars with an optically identified extragalactic
population of $\approx$ 13\% (Guillout 1996).  At higher galactic latitudes, the fraction
of active coronae identified in the RASS falls to $\approx$ 42\% (Zickgraf et al. 1996) and
may reach values as low as 10\% at the north galactic pole (Zickgraf 1996).  It is thus
amazing that in spite of the apparently nearly constant density of ROSAT survey sources
with galactic latitude ($\approx$ 1.5 deg$^{-2}$; Voges 1992) their nature changes from an
extragalactic dominated to star dominated population while moving to lower galactic
latitudes.  

\begin{table*}

\caption{Source identification statistics in the 'full' area (all ML). Figures are number
of sources}

\label{overallstat1}

\begin{tabular}{lccccccc}
Limiting         & Total number & Active    & OB    &  AGN  & CV or   & White  & Unidentified \\
count rate       & of sources& coronae   &       &       & related & dwarfs  &               \\
\hline
               0.1  &  12    &  8        & -     &  1    &  -      & 3      & -  \\
               0.05 &  21    &  17       & -     &  1    &  -      & 3      & -  \\
               0.03 &  48    &  41       & 2     &  1    & -       & 3      & 1  \\
               0.025&  57    &  48       & 2     &  1    & -       & 3      & 3  \\
               0.02 &  68    &  54       & 2     &  1    & -       & 3      & 8  \\
               0.015& 101    &  69       & 2     &  1    &  1      & 3      & 25 \\ 
               0.01 & 138    &  77       & 3     &  1    &  2      & 3      & 52 \\
\hline

\end{tabular}
\end{table*}

\begin{table}
\caption{Source identification statistics in the 'full' area. Figures are in percents}
\label{overallstat2}
\begin{tabular}{lrr}
Limiting         & Fraction of  &   Fraction      \\
Count rate       & Active Coronae     &   Identified  \\
\hline
            0.1  &  67                &   100 \\
            0.05 &  81                &   100 \\
            0.03 &  85                &   98  \\
            0.025&  84                &   95  \\ 
            0.02 &  79                &   88  \\
            0.015&  68                &   75  \\
            0.01 &  56                &   62  \\
\hline

\end{tabular}
\end{table}

\subsection{Positions and 90\% confidence radii}

An accurate knowledge of the ROSAT error circle is of utmost importance for guiding the
optical observations in the crowded fields often encountered at low galactic latitude. An
illustrative example is probably our discovery of the WD+Me binary RX\,J2130.3+4709 (index
84).  Spectroscopy of the bright (V = 8.45) G0V star HD 204906 failed to reveal convincing
\ion{Ca}{II} emission and furthermore the star was clearly outside the \rqvd \ radius. This
led us to investigate other candidate stars closer to the X-ray position and discover this
interesting system.  

\begin{figure}
\psfig{figure=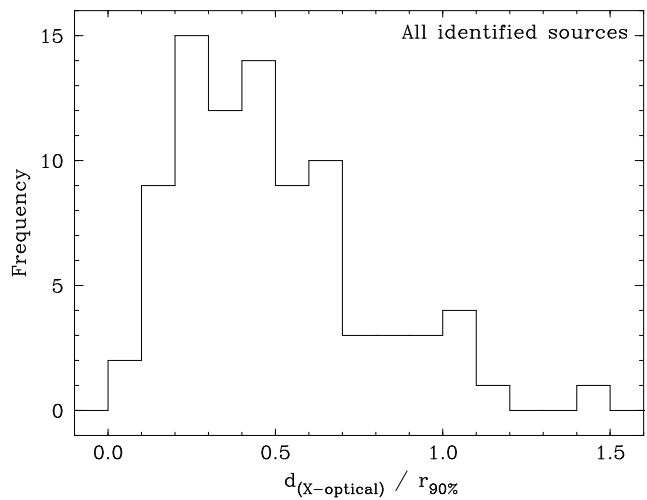,width=8.8cm,bbllx=2cm,bburx=20cm,bblly=2cm
,bbury=17cm}

\caption[]{Histogramme of the distances between X-ray and optical positions expressed in
units of the computed 90\% confidence radius. The distribution is shown here for the
86 sources identified in the Cygnus area}

\label{dxo_r90}   
\end{figure}

The errors on ROSAT X-ray positions have two different origins. First, the uncertainty
with which the centroid of the X-ray image is positioned on the pixel grid by the
maximum likelihood source detection algorithm. Second the error in the knowledge of the
satellite attitude for each photon collected in scan mode. Early analysis led to the
conclusion that the error on the position was dominated by the ROSAT attitude bore
sight accuracy and that assuming a Gaussian two-dimensional distribution the final
uncertainty could be written as:

\begin{displaymath} 
{\rm r}_{90}  \ = \ 2.15 \times \sqrt{ \Delta^{2}_{xy} \ +\Delta^{2}_{att} } 
\end{displaymath}

\noindent 

where $\Delta_{xy}$ is the maximum likelihood error and where $\Delta_{att}$, the attitude
uncertainty was estimated to be $\approx$ 8\arcsec \ from a subset of X-ray binaries with
accurate positions (Motch et al. 1996b).  Applying this estimate to the sources in our
survey field yields a mean \rqvd \ in the range of 19\arcsec \ to 32\arcsec \ slightly
depending on the maximum likelihood of detection (see Table \ref{r90stat}).  The histogram
of the distance between X-ray and optical positions expressed in units of the 90\%
confidence radius (see Fig. \ref{dxo_r90}) is compatible with a normal Rayleigh
distribution. The fact that only $\approx$ 7\% of all 86 proposed optical counterparts are
located outside \rqvd \ from the ROSAT position suggests that we may have slightly
overestimated the satellite attitude error and that the actual $\Delta_{att}$ is close to
7\arcsec. Adding the 14 less secure candidates identified at the 95\% confidence level only
does not change this conclusion. However, the identification strategy may have favoured
optical counterparts located close to the X-ray position. The positions extracted from
SIMBAD are based on 1950 coordinates and are not corrected for proper motion. Also the
"Quick V" plate collection on which are based the northern GSC coordinates was obtained in
the 1982-1984 interval and these positions may be altered to some extent by the unknown
proper motion. Therefore, a small fraction of the scatter in the distance between X-ray and
optical positions may be due the lack of correction for proper motion.

\begin{table}
\caption{90\% confidence radii for various maximum likelihood}
\label{r90stat}
\begin{tabular}{ccc}
Maximum likelihood   & Mean            & Mean\\
of detection         & \rqvd (\arcsec) & Count rate (\cnts )\\
\hline
$\geq$ 100  & 19.2 & 0.300\\
50 - 100    & 22.4 & 0.066\\
20 - 50     & 23.8 & 0.035\\
10 - 20     & 28.8 & 0.019\\
$\leq$ 10   & 32.3 & 0.012\\
\hline
\end{tabular}
\end{table}

\subsection{X-ray hardness ratios}

\begin{figure}
\psfig{figure=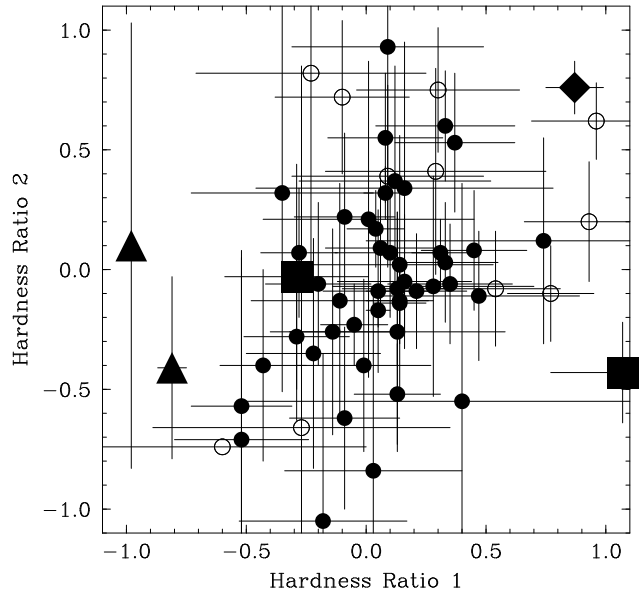,width=8.8cm,bbllx=1cm,bburx=18cm,bblly=2cm
,bbury=19cm}

\caption[]{Distribution of a subset of 62 sources having computed hardness ratios.
Unidentified sources are represented as open circles.  Identified active coronae
(filled circle) populate the center of the diagram and span a rather large range in
HR1 and  HR2. White dwarfs (filled triangle) with soft HR1 clearly stand out of the
dominant stellar population. The only optically identified AGN (filled diamond) is
hard in both X-ray colours as a result of the large intervening photoelectric
absorption. Of the two objects flagged as 'OB' stars (filled squares) one is a
B9III/Ap candidate which displays X-ray colours similar to active coronae. The early
type star HR 8154 exhibits very hard HR1 due to interstellar absorption and rather
soft HR2 consistent with the emission of an O8Ve star at the distance of the OB
association CYG OB7}

\label{hr1hr2}   
\end{figure}

\begin{figure}
\psfig{figure=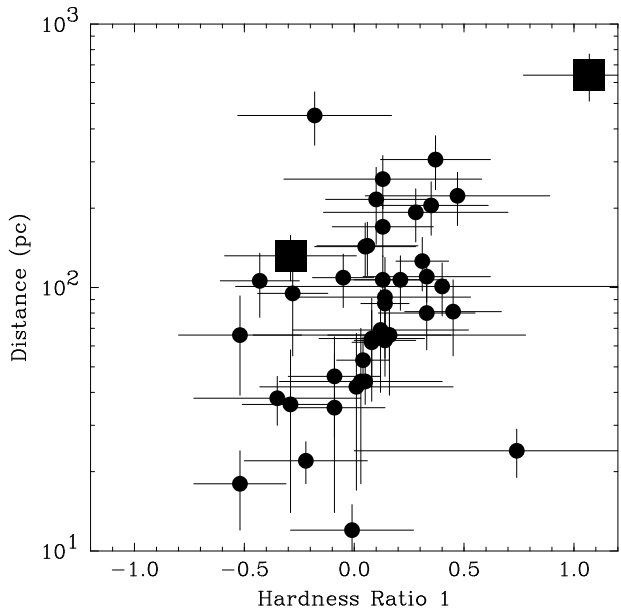,width=8.8cm,bbllx=1cm,bburx=18cm,bblly=2cm
,bbury=19cm}

\caption[]{Distribution in HR1 and distances for a subset of active coronae (filled
circles) and OB and B9III/Ap stars (filled squares). Distant sources may display
harder X-ray spectra than nearby ones. In the absence of detailed spectral fitting
this effect could be either due to interstellar photoelectric absorption and / or to
the higher temperatures usually observed in the most luminous active coronae} 

\label{hrdist} 
\end{figure}

In principle hardness ratios are powerful tools to select and guide the identification of
ROSAT X-ray sources (see e.g. Motch 1992). However, the use of X-ray colours is essentially
limited to the brightest sources. With a mean exposure time of the order of 800\,s useful
information may be extracted for the sources having count rates larger than $\approx$
0.02\,\cnts . We show in Fig. \ref{hr1hr2} the distribution of a subset of 62 sources with
count rates larger than 0.012\,\cnts, in the HR1/ HR2 diagram.  Hardness ratios 1
and 2 have here energy boundaries as used in the latest SASS versions:

\begin{displaymath}
{\rm HR1}\ = \ \frac{(0.5-2.0)-(0.1-0.4)}{(0.1-0.4)+(0.5-2.0)}
\end{displaymath}
\begin{displaymath}
{\rm HR2}\ = \frac{(1.0-2.0)-(0.5-1.0)}{(1.0-2.0)}
\end{displaymath}

\noindent where (A-B) is the raw background corrected source count rate in the A-B energy
range expressed in keV.  Identified active coronae span a wide range of X-ray colours
consistent with the variety of temperatures encountered in stars (e.g. Schmitt et al. 
1990).  White dwarfs populate the soft region of the diagram while our only identified AGN
exhibits the hard X-ray colours expected from a heavily absorbed source. The unidentified
source RX\,J2133.3+4726 (index 35) is located on the sky close to the identified AGN and
displays hard X-ray colours (HR1 = 0.96 $\pm$ 0.27; HR2 = 0.62 $\pm$ 0.16). Deep optical
searches in this error circle indeed rule out identification with active coronae (Motch et
al. 1996a). This suggests a remote luminous source possibly of extragalactic nature. The
O8Ve star HR 8154 also exhibits a soft HR2 and hard HR1 which probably originates from a
substantial interstellar absorption toward this remote (d $\approx$ 640\,pc) source in CYG
OB7.  Active coronae detected at large distance may have harder HR1 than closer ones (see
Fig.  \ref{hrdist}). However, this effect could either be due to enhanced photoelectric
absorption towards most distant objects and / or to intrinsically higher temperatures
usually observed for the most luminous active coronae which are obviously the ones we detect
at the largest distances (Schmitt et al. 1990).  

\subsection{X-ray variability}

\begin{figure}
\psfig{figure=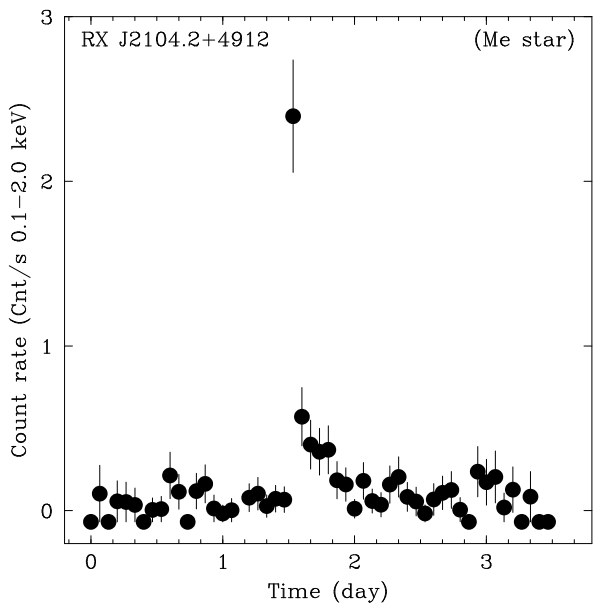,width=8.8cm,bbllx=1cm,bburx=18cm,bblly=2cm
,bbury=19cm}

\caption[]{X-ray flare from the Me star RX\,J2104.2+4912 (index 7). Few sources exhibit 
flaring-like activity and we conclude that there is probably no large bias favouring the
preferential detection of particularly active stars}

\label{xflare} 
\end{figure}

Owing to the snapshot nature of survey observations for which we have at best a 32\,s
long integration every 96\,min only limited information on time variability is
available. However, active coronae are known to exhibit large X-ray flares and one may
wonder whether a sizeable fraction of our sources is detected thanks to these fast
events. Visual inspection of the light curves produced in the broad (0.1-2.0\,keV) and
hard (0.5-2.0\,keV) X-ray band reveals that most active coronae were detected during
all satellite orbits and that the flaring activity is in general not prominent. Our
only conspicuous flare originates from the Me star RX\,J2104.1+4912 (index 7; see Fig.
\ref{xflare}). In the absence of this particular event, the mean count rate from the
source would have been 0.062\,\cnts \ instead of the 0.13\,\cnts . Accordingly, we conclude
that the effect of flares is not important in our sample and that there is no large
bias in favour of the preferential detection of flare stars.

\section{Stellar population}

As stated above, the population of active coronae dominates our sample and down to a
flux level of 0.03\,\cnts \ stars account for 85\% of the total number of field
sources. Unfortunately, we do not have spectra for all star candidates in the 'full'
area. However, in the restricted 'inner' area (38.1\,deg$^{2}$) we have spectroscopic
information from the literature or from our own observations for all 42 identified
X-ray stars detected above a count rate of 0.02\,\cnts .  We list in Table
\ref{spstat} the repartition in spectral type of this complete subsample.  

\begin{table}

\caption{Repartition in spectral types of all X-ray stars (giant and main sequence) with
count rate $\geq$ 0.02\,\cnts \ and located in the restricted 'inner' area}

\label{spstat}
\begin{tabular}{crc}
Spectral    & Numbers  & Percentages         \\
Type        &          &                     \\
\hline
A           & 6        & 14.3 $\pm$ 5.4 \\
F           & 7        & 16.7 $\pm$ 5.7 \\
G           & 11       & 26.2 $\pm$ 6.8 \\
K           & 10       & 23.8 $\pm$ 6.5 \\ 
M           & 8        & 19.0 $\pm$ 6.0 \\
Total       & 42       & 100  \\     
\hline
\end{tabular}
\end{table}

\begin{figure}
\psfig{figure=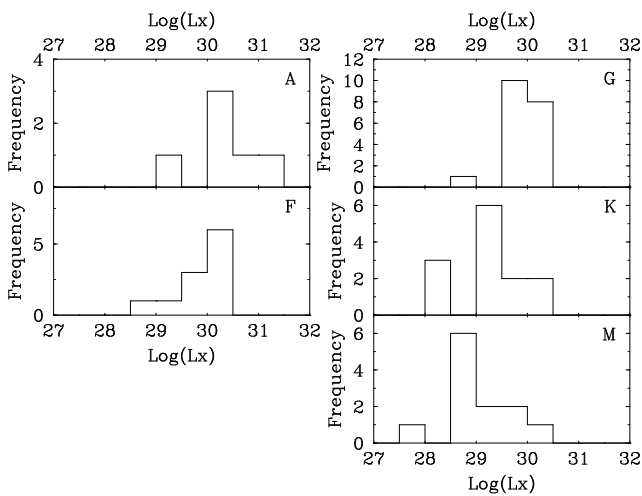,width=8.8cm,bbllx=3cm,bburx=23cm,bblly=2cm
,bbury=18cm}

\caption[]{Distribution in X-ray luminosities of all identified active coronae having
spectral type information either from the literature or from our own optical
observations. X-ray luminosities were computed assuming negligible interstellar
absorption and a (0.1-2.4\,keV) count to flux conversion factor of 1.0 $\times$
10$^{-11}$ erg cm$^{2}$ s$^{-1}$.  As expected in an X-ray flux limited sample we
preferentially sample the high luminosity tail of the X-ray luminosity function} 

\label{lxhisto} 
\end{figure}

It is known that single A stars are very weak X-ray emitters in agreement with their
almost absent convective zone and chromospheric activity.  Therefore our identification of
a small fraction of X-ray sources with A stars, here only based on a very low probability
of random coincidence, may only be understood if one assumes the presence of an X-ray
active companion of lower mass. Considering the rather short main sequence life time of A
stars ($\leq$ 10$^{9}$ yr) any coeval companion star will probably have a high X-ray
activity as a result of its young age. However, ROSAT HRI observations seem to question
this explanation and may point toward intrinsic X-ray emission in some late B stars
(Bergh\"ofer \& Schmitt 1994). At the distance of the A stars the observed X-ray
luminosities (see Fig.  \ref{lxhisto}) are well in the range of those observed for
classical active coronae. In two cases (RX\,J2056.7+4940, index 20 and RX\,J2117.8+5112,
index 34) we obtained multicolour CCD images of the field but failed to identify a nearby
likely alternative candidate. Assuming a limiting angular distance of 1\arcsec \ and a
mean distance of 250\,pc implies an upper limit of $\approx$ 250\,AU on the projected
separation of the two stars.  

Only two giant stars HR 8252 (RX\,J2133.9+4535, index 2) and HR 8072 (RX\,J2103.4+5021, index 37) 
are present in our spectroscopically identified sample in agreement with the
low spatial density and moderate X-ray luminosities of the evolved late type
stars (Maggio et al. 1990). 

The number of identified close systems is also small. We find evidences for a short period
binary in two cases, RX\,J2107.3+5202 (index 26, V1061 Cyg; eclipsing F8+G1, P =
2.3467\,d, Dworak 1976) and RX\,J2035.9+4900 (index 53, exhibiting a clear line doubling,
Motch et al. 1996a).  In a subsequent article we shall discuss in more detail the
contribution of close binaries using as additional material red optical spectra covering
the Li $\lambda$ 6707 \AA \ line at a slightly higher resolution than that of the blue
medium resolution spectra discussed here. In one additional case, RX\,J2120.9+4636 (index
9) we find conspicuous line doubling in the red medium resolution spectra indicating a
close binary.

Our stellar X-ray population displays all the characteristics of an X-ray flux limited
sample. We preferentially detect the high X-ray luminosity tail of the distribution
function, several stars exhibiting X-ray luminosities in excess of 10$^{30}$ \ergs  \
(see Fig. \ref{lxhisto}). X-ray luminosities were computed assuming negligible
interstellar absorption and their estimated error is of the order of $\approx$ 40\%
(Motch et al. 1996a).  The mean X-ray luminosity clearly decreases with cooler
photospheric effective temperature. This effect could be the result of a
dependence of X-ray luminosity with stellar radius (Fleming et al. 1989). We note that
the only M star observed above 10$^{30}$ \ergs \ is the flaring source RX\,J2104.1+4912
(index 7, see Fig. \ref{xflare}) which would have been detected at about half of this luminosity
in the absence of the flare.  No particular spectral type dominates the stellar content
of our survey area.  Although the number density of dwarf G/K stars is only 10 to 16\%
that of dwarf M stars our distribution peaks at the G/K types.  These number ratios
indicate that the detection volume for G and K stars are about 14 and 8 times that of M
stars and that accordingly, the mean X-ray luminosities of G and K stars are 5.7 and 4
times that of M stars. These figures are in general in good agreement with the mean
Einstein X-ray luminosity ratios for disk stars (Micela et al. 1988, Barbera et al.
1993).

\begin{figure}
\psfig{figure=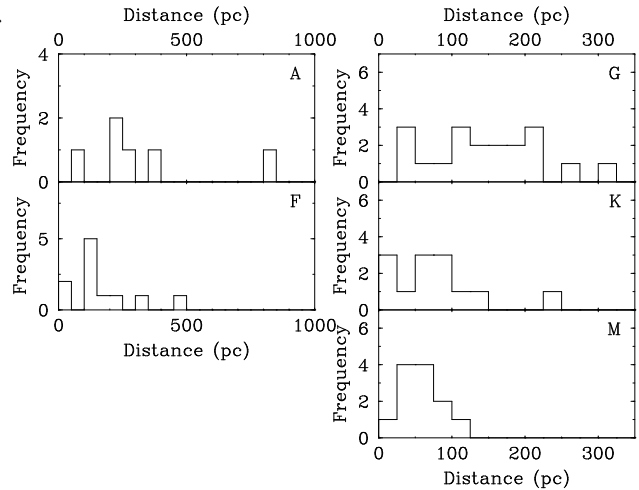,width=8.8cm,bbllx=3cm,bburx=23cm,bblly=2cm
,bbury=18cm}
\caption[]{Distribution in distance of all identified active coronae having spectral
type information either from the literature or from our own optical observations}
\label{distancehisto} 
\end{figure}

\begin{figure}
\psfig{figure=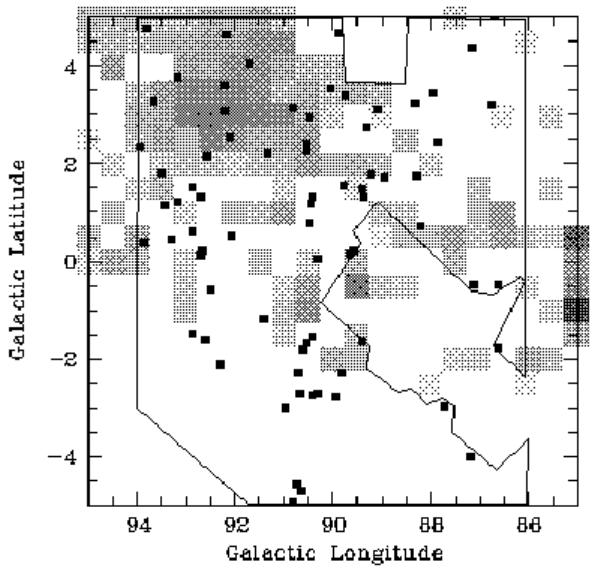,width=8.8cm,bbllx=5cm,bburx=16cm,bblly=8cm,bbury=20cm}

\caption[]{Position of the identified active coronae (maximum likelihood $\geq$ 8)
overlayed on the CO map from Dame et al. (1987) where we also show the area
investigated in X-rays. Among the 24 open clusters present in a 11\degree
$\times$11\degree \ region centered at $l$ = 90\degree, $b$ = 0\degree, only M39
(d $\approx$ 300\,pc) at $l$ = 92.5\degree , $b$ = -2.3\degree \ is close enough to be
possibly detected at our sensitivity. The absence of concentration of sources at this
position indicates that the presence of the cluster does not significantly bias the
stellar statistics}

\label{acidco}   
\end{figure}

Obviously, our survey maximum sampling distance also depends on the spectral type and
we detect F stars up to 500\,pc and M stars only up to $\approx$ 100\,pc (see Fig. 
\ref{distancehisto}). Among the 24 open clusters recorded in SIMBAD in a 11\degree
$\times$ 11\degree \ region centered at $l$ = 90\degree, $b$ = 0\degree , only M39 at
($l$ = 92.5\degree , $b$ = -2.3 \degree ) is close enough (d = 300 $\pm$ 30\,pc; Mohan
\& Sagar 1985) to possibly introduce some inhomogeneity in our X-ray stellar sample.
The relatively young age of the cluster (3$\pm$1 10$^{8}$ yr, Mohan \& Sagar 1985)
indeed suggests that some active coronae could be detected.  However, we do not
observe any strong clustering of identified active coronae (see Fig. \ref{acidco})
nor unidentified sources in that particular direction (see Fig. \ref{unidco}). Among
the 4 X-ray stars detected within 1.5\degree \ from the center of the open cluster,
only RX\,J2130.7+4919 (index 110) identified with Star 1930 in the Platais (1994)
catalogue could be member of M39.  Therefore, we believe that as far as active coronae
are concerned our sample area is indeed representative of average mid galactic plane
conditions.  

\section{Accreting and related sources}

Compared to other regions in the galactic plane this area in Cygnus is particularly
void of accreting sources. The only catalogued X-ray binary is the low mass system 4U
2129+47 at $l$ = 91.58\degree \ and $b$ = -3.04\degree \ which is presently in an
extended low state since the early 1980s (Pietsch et al. 1986).  Our non detection is
consistent with the faint ROSAT HRI source detected with a count rate of 0.004\,\cnts
\ during a pointed observation which took place about one year after the survey
observations (Garcia 1994).  The only cataclysmic variable appearing in our source
list is V1500 Cyg, also known as Nova Cyg 75, which is detected at the rather low 
maximum likelihood of 7.6. 

We identify  RX\,J2130.3+4709 (index 84) with a WD+Me star close binary. The blue
optical spectrum (see Motch et al. 1996a) displays heavily broadened Balmer lines
typical of a rather cool white dwarf with superimposed narrow Balmer and \ion{Ca}{II} H\&K
emission lines.  In the red, the spectrum is dominated by that of a late Me star with
strong TiO molecular bands and \Halp \ emission.  Radial velocities of Balmer
absorption and emission lines vary in opposite directions with an orbital period of
$\approx$ 12 hours. The white dwarf in RX\,J2130.3+4709 is too cool to contribute to
the X-ray emission and the flux detected by ROSAT (1.6 10$^{-2}$ \cnts ) probably
arises from the Me star companion. In such a compact binary the late type star may be
tidally locked to the white dwarf implying an enhanced rotational velocity and
resulting coronal activity. RX\,J2130.3+4709 appears quite similar to RX\,J0458.9-6628
(Hutchings et al.  1995). We shall extensively discuss the properties of this system
and its possible relation to cataclysmic variables in a forthcoming paper.
 
\section{Extragalactic contamination}

Although very low galactic fields such as the one investigated here are unlikely to
contain a large fraction of extragalactic sources, our discovered AGN at $b$ =
-3.4\degree \ shows that this population is not completely screened out and may account
for a significant fraction of the optically unidentified and relatively faint X-ray
sources.  In order to estimate this 'pollution' we assumed that the un-absorbed
extragalactic \lnls \ function was essentially that given in Hasinger et al. (1993).
At our maximum possible flux sensitivity of $\approx$ 1-2 10$^{-13}$ \ergcm \ we are
far above the departure from the near Euclidean distribution and we may write the number of
sources having X-ray flux within $s$ and $s\ + \ ds$ as: \begin{displaymath} n(s) \ ds \ =
\ N \, s^{-\beta} \ ds \end{displaymath}

However, our only measured quantity is the source count rate and for a given intrinsic
spectrum the flux over count ratio may depend heavily on the intervening photoelectric
absorption. In the Lockmann Hole (Hasinger et al. 1993) the soft X-ray flux is $s$ =
$k_{0} \, \times \, c$ where $c$ is the count rate measured from the source and the
differential extragalactic count distribution is:

\begin{displaymath}
n(c) \ = \ N \, c^{-\beta} k_{0}^{1-\beta}
\end{displaymath}

The observed count rate from sources dimmed by galactic absorption may be expressed as
$c_{a} \ = \ K(\Omega) \, c$ where $K(\Omega)$ is the count absorption coefficient at
position $\Omega$. Averaging over the whole area and using the explicit form of the
differential count distribution we may write:

\begin{displaymath}
n(c_{a}) \ = \frac{N\,c_{a}^{-\beta}\,k_{0}^{1-\beta}}{\Omega_{0}} \, 
\int_{\Omega_{0}} K(\Omega)^{\beta} \, d\Omega
\end{displaymath}

Therefore, the absorbed extragalactic distribution has the same slope as the
un-absorbed one but shifted to lower count rates.  

We assumed a \lnls \ (0.5-2\,keV) relation with $N$ = 104, $\beta$ = 2.44 and an
average spectrum of extragalactic sources in the form of a power law of energy index
0.96 (Hasinger et al. 1993, Gioia et al. 1990).  We computed the count to flux
ratios and count absorption coefficients for different values of the galactic
absorption by folding the incident absorbed spectra with the detector response curve
and by fitting a three degree polynomial to the log(K)/log(\nh ) relation.  Because of
the high photoelectric absorption we neglected the count rate resulting from the
0.1-0.5\,keV low energy part of the spectrum.  The galactic absorption was estimated
from the HI maps of Strong et al. (1988) and Dame et al. (1987) interpolated on a
0.5\degr \  grid. The total \nh \ was taken as N$_{\rm HI}$ + 2 $\times$ 3.5 10$^{20}$
$\times$ T$_{\rm CO}$ (e.g. Solomon \& Rivolo 1987). Finally, the K coefficient was
then integrated over the observed area.  

Our simulation shows that on the average Cygnus field the count rate of extragalactic
sources is only 15\% of that detected at high galactic latitude and that at a given
flux level we only detect 3.5\% of the total extragalactic population.  Table
\ref{extralnls} lists the expected number of extragalactic sources as function of the
absorbed count rates.  
 
\begin{table}
\caption{Absorbed extragalactic \lnls \ function}
\label{extralnls}

\begin{tabular}{rrr}
  count rate $ c_{a} $        & N ( $\geq \ c_{a}$ ) &  N ( $\geq \ c_{a}$ )    \\
  0.5-2\,keV cnt s$^{-1}$      &   1 deg$^{2}$        &   'full' area      \\                   
\hline
  0.01                        &    7.3 10$^{-2}$     &  4.7      \\
  0.02                        &    2.7 10$^{-2}$     &  1.7      \\
  0.03                        &    1.5 10$^{-2}$     &  1.0      \\
  0.05                        &    7.1 10$^{-3}$     &  0.45      \\
  0.10                        &    2.7 10$^{-3}$     &  0.17      \\
  0.50                        &    2.6 10$^{-4}$     &  0.02      \\
\hline
\end{tabular}
\end{table}

\begin{figure}
\psfig{figure=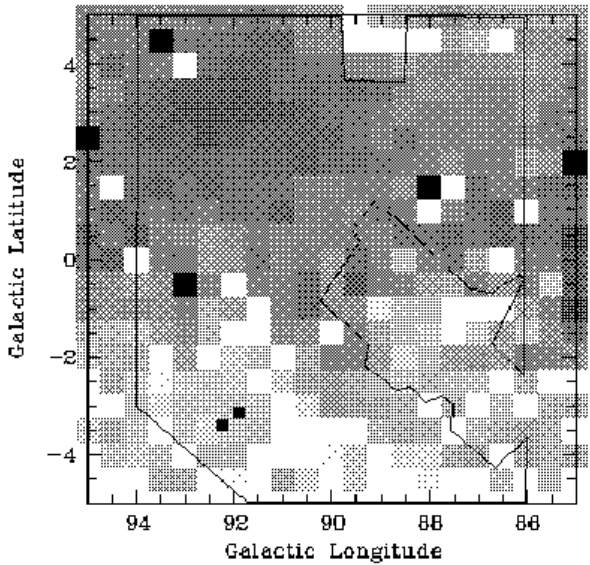,width=8.8cm,bbllx=5cm,bburx=16cm,bblly=8cm,bbury=20cm}

\caption[]{A total \nh \ (HI + H$_{2}$) map (log grey scale) where we also show the
region investigated in X-rays. \nh \ varies between 4.4 10$^{21}$\,cm$^{-2}$ up to 4
10$^{22}$\,cm$^{-2}$. Not surprisingly, our two extragalactic candidates (filled
squares) fall in regions of low galactic absorption.  We estimate that over the whole
field shown here the extragalactic sources do not account for more than a few percent
of the total population detected at a count rate above 0.02\,\cnts }

\label{fig_extragalactic}   
\end{figure}

We have only one positive optical identification with a Seyfert 1 nucleus at a count rate of
0.1\,\cnts \ (RX\,J2135.9+4728, index 12). Another source (RX\,J2133.3+4762, index 35) with a
count rate of 0.035\,\cnts \  is likely extragalactic. It has a very hard spectrum similar to
that of the identified AGN and deep optical investigations fail to reveal the counterpart. We
show in Fig. \ref{fig_extragalactic} the position of our two extragalactic candidates on a \nh \
map together with the investigated area.  It is probably meaningful that our two non-galactic
candidates fall into a region of very low absorption.  

Considering the small number statistics, we believe that the number
and count rates of our identified extragalactic candidates (2 sources with $ c_{a} \
\geq $ 0.03\,\cnts ) are in accord with our simulation. We conclude that the
extragalactic population is likely to represent only a very small fraction ( $\approx$
3\%) of the 68 sources detected in the 'full' area above our completeness level of
0.02\,\cnts .  

\section{Discussion}

\subsection{The \lnls \ relation}

\begin{figure}
\psfig{figure=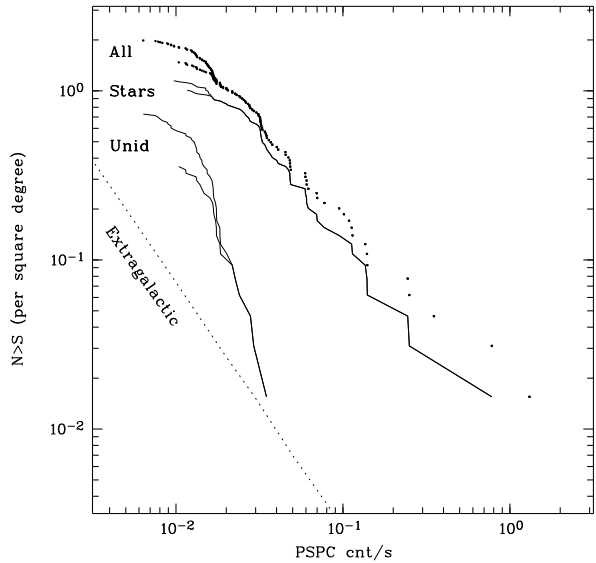,width=8.8cm,bbllx=0cm,bburx=21cm,bblly=0cm,bbury=20cm}

\caption[]{The \lnls \ function for sources with ML $\geq$ 10 (lower curves) and ML
$\geq$ 8 (upper curves) in the 'full' area (64.5 square degrees).  Above the
completeness level of $\approx$ 0.02\,\cnts \ the slope of the distribution is close to
-1. The identified stellar component has a slope of -1.48 $\pm$ 0.23 consistent with
the Euclidean value for count rates above our identification completeness level of
$\approx$ 0.03\,\cnts .  The dashed line represent the expected extragalactic
contribution} 

\label{lognlogs}   
\end{figure}

We show in Fig. \ref{lognlogs} the number count relation for all sources detected in
the 'full' area above ML = 8 and ML = 10. From the bending of the \lnls \ curve at low
count rates we estimate that our total source sample is complete down to about 0.02
and 0.012 \cnts \ for ML = 10 and 8 respectively. We note, however, that the enhanced
number of spurious sources (see section 9.3) may complicate the definition of a 
completeness level at ML = 8. The corresponding source densities are 1.0 deg$^{-2}$
and 1.7 deg$^{-2}$ at 0.02 and 0.012 \cnts \ respectively.  In order to estimate the
slope of the overall relation we used the maximum likelihood technique of Crawford et
al. (1970) and Murdoch et al. (1973).  For the 67 sources with ML $\geq$ 10 and count
rate $\geq$ 0.02 \cnts , we find that: 

\begin{displaymath} 
N ( \geq S ) = 1.7\times 10^{-2} \ S^{-1.05\pm0.13} 
\end{displaymath}

The slope and normalisation of the ROSAT survey relation is fully consistent with that
reported by Hertz and Grindlay (1984) for the Einstein Galactic Plane Survey. Morley et
al. (1996) also report compatible \lnls \ relations from deep pointings in the galactic
plane.  As discussed in the previous section, the extragalactic contribution is
probably negligible in our field.  At high count rates the relative excess of sources
is caused by few objects namely the three white dwarfs and our only identified AGN. In
principle the slope of the white dwarf \lnls \ function should be close to -1.5. In
fact the lack of identified white dwarfs with count rates below 0.11\,\cnts \ could
reflect the higher completeness flux level for this particular population resulting
from the higher X-ray background at low energies.  Alternatively we may see the large
effects of tenuous interstellar absorption on such soft X-ray spectra.  

Our high fraction of identified sources put us in a position to derive the \lnls \ relation
for stellar sources independently of the overall population. The inflexion of the \lnls \
curve for the stellar X-ray sources also shown in Fig.  \ref{lognlogs} suggests that the
identified stellar installment is roughly complete down to $\approx$ 0.03 \cnts \ in the
'full' area, while the completeness level reaches 0.02 \cnts \ in the 'inner' area (see
sect. 6).  Interestingly, we find that the \lnls \ function of the optically identified
active coronae is steeper than that of the total X-ray population. For the 41 sources with
ML $\geq$ 10 and count rates above our estimated completeness level of identification
($\approx$ 0.03 \cnts ) we find a slope consistent with the Euclidean value.

\begin{displaymath}
N_{\rm stars} ( \geq S ) = 3.6\times 10^{-3} \ S^{-1.48\pm0.23}
\end{displaymath}

This indicates that at the flux level of the ROSAT all-sky survey the distribution of
the detected stellar population is not severely affected by interstellar absorption. 
The Euclidean-like distribution is also consistent with the fact that in our low
latitude test direction, the maximum distance above the plane ($\approx$ 30\,pc) of
the most remote X-ray detections ($\approx$ 300\,pc) is much smaller than the scale
height of any known group of late type stars.

\subsection{X-ray counts modelling}

\begin{figure*}[t]
\psfig{figure=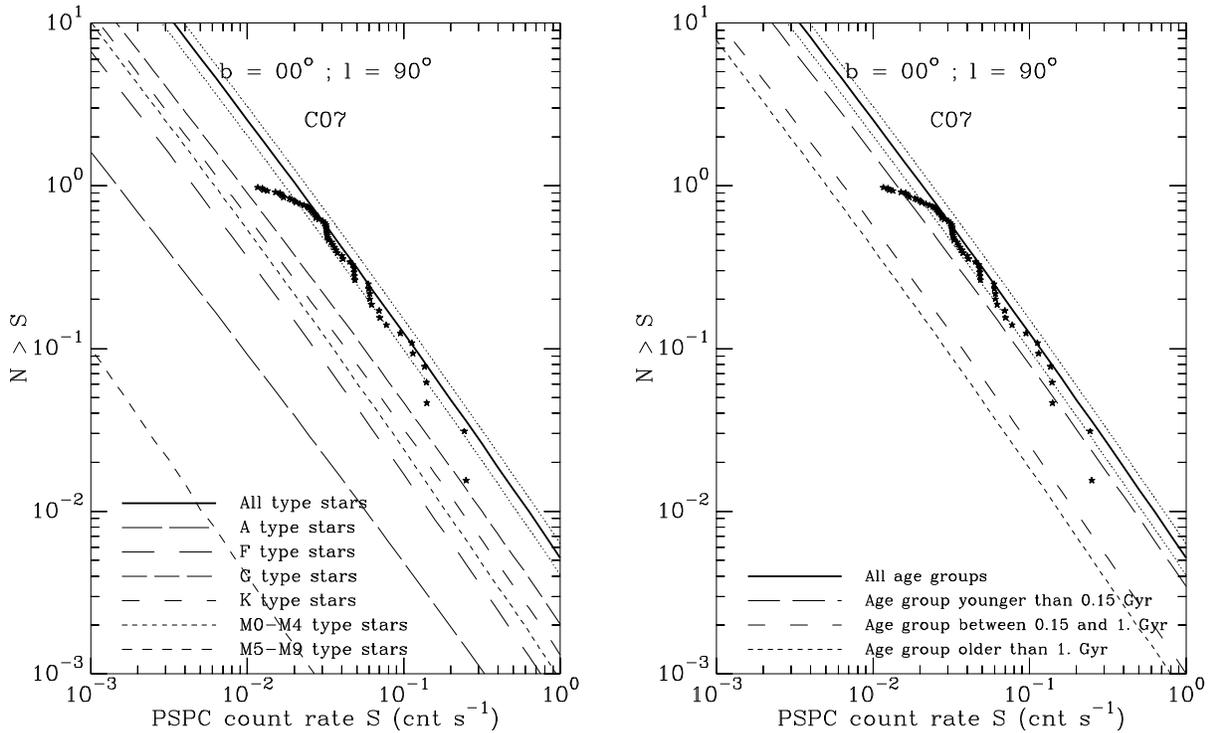,width=18cm,bbllx=0.2cm,bburx=20.8cm,bblly=8cm,bbury=21.5cm}

\caption[]{Theoretical \lnls \ curves for all stellar coronae (thick line), computed
for the direction of the test area, assuming a constant stellar formation rate and a
slope of the initial mass function below 1 $M_{\odot}$ equal to 0.7 (C07).  N is the
number of stars per square degree in the direction $l$,$b$ up to a given PSPC count rate S
(0.1 - 2.4\,keV band) as a function of the count rate. We also show the \lnls \ curves
computed separately for A,F,G,K and M type stars (a) - left panel) and various
populations of disc stars (b) - right panel). The thin lines on both sides of the sum
\lnls \ curves represent the possible range resulting from the error due to the
binning in age of the X-ray luminosity functions. The observed relation for all
identified active coronae (giants excluded) with ML $\geq$ 10 in the 'full' area is
shown with asterisks}

\label{popmodel}
\end{figure*}

Optical star count modelling is a very powerful mean to constrain several important
parameters describing the various stellar populations of the Galaxy (e.g. Robin \&
Cr\'ez\'e 1986).  Because of the strong dependence of stellar X-ray emission with age, an X-ray
view of the sky preferentially reveals the young stellar population.  X-ray count
modelling such as that developed by Favata et al. (1992) allows in particular the
study of the local star forming rate during the last 10$^{9}$~yr, a field of
investigation which is usually loosely constrained by optical star count analysis. 
Looking deep into the galactic plane has specific advantages since it allows to better
constrain the youngest most luminous population which is concentrated in the low
latitude regions of the galactic plane.  

We first tried to compare our density of active coronae with the predictions of Favata
et al. (1992).  In their low galactic latitude model ($b$ = 10\degree ) the number of
X-ray stellar source counts above log(F$_{X}$/erg cm$^{-2}$ s$^{-1}$) = -12.5
corresponding to a PSPC count rate of $\approx$ 0.03\,\cnts \ is $\sim$ 0.1 per square
degree while our observed density is $\approx$ 0.6. This rather large discrepancy
could be related to the way the youngest stellar populations are treated.  

In a second step we designed an age dependent numerical model by folding X-ray
luminosity functions (XLFs) with the stellar population model of Robin \& Cr\'ez\'e
(1986). We used XLFs derived from ROSAT observations of the Pleiades and Hyades young
galactic clusters and from Einstein observations of old disc population to handle the
steeply varying stellar X-ray activity with age. Binaries are accounted for by using
binary corrected luminosity functions. This X-ray model (fully described in Guillout
et al. 1996a) allows to predict the distribution in age, spectral type, various colour
indices, magnitudes and distance of the stellar content of X-ray flux limited or
volume limited surveys. Detailed comparison of model predictions with observations in
several low galactic latitude RASS areas will be discussed elsewhere and we only
present here a preliminary analysis. Fig.  \ref{popmodel} shows the computed \lnls \ 
curves toward the direction of observation ($l$ = 90\degree , $b$ = 0\degree) for the
standard model (stellar formation rate constant, slope of the initial mass function
below 1 $M_{\odot}$ = 0.7 $\pm$ 0.2) (Haywood 1995).  The most interesting feature in
Fig. \ref{popmodel} is the good agreement between observations and the predicted
number counts curve. The departure at count rates below 0.02 \cnts \ is probably due
to the incompleteness of our optical identifications. This suggests that in this
sample region we clearly detect all the active stars expected on the basis of our
current knowledge on stellar population and age dependent X-ray coronal activity.  

Table \ref{poptable} summarizes the model number of active coronae and the proportion
of A, F, G, K, early and late M-type stars for various limiting PSPC count rate (0.1 -
2.4\,keV band). For comparison, we list in Table \ref{poptable2} the distribution in
spectral types of a completely identified stellar subsample made of the 40 main sequence
stars located in the 'inner' area with S $\ga$ 0.02\,\cnts \ together with the predicted
distribution at the same flux limit.  Considering the small number statistics, the
observed distribution is in reasonable agreement with model predictions.

Another important feature shown in Fig.  \ref{popmodel} is the outstanding importance of
very young stars in the RGPS.  For a limiting PSPC count rate S = 0.03\,\cnts \ we
expect that $\approx$ 65\% of the detected stellar X-ray sources are younger than 0.15
Gyr, this proportion increasing up to $\approx$ 85\% for stars younger than 1 Gyr (see
Table \ref{poptable}).  These first results indicate that stellar X-ray source counts
may indeed be used to derive the basic properties of the young stellar populations and
in particular put interesting constraints on the stellar formation rate (e.g. Micela et
al. 1993, Guillout et al. 1996b).

The X-ray count model used here ignores the contribution of giant stars which is
expected to be small (Maggio et al. 1990). Another population not taken into account is
that of close 'old' binaries. In these systems, the large rotation velocity maintained
over the whole stellar life by tidal lock may sometimes yield large X-ray luminosity. 
However, the large uncertainties on the X-ray luminosity functions, spatial densities,
mass ratio distributions of close binaries such as RS CVn systems (e.g.  Ottmann \&
Schmitt 1992, Favata et al. 1995) do not allow to accurately estimate their specific
contribution. Using the XLFs in Majer et al. (1986) and Favata et al. (1995), we
estimate that more than 3\% of the sources brighter than 0.02 \cnts \ could be active
close binaries which is close to the observed fraction (4\%) in our sample. However, as
discussed in section 6, our follow-up observations are not very sensitive to binarity
and the actual number of close systems may be somewhat larger.  

\begin{table*}

\caption{Predicted number of stellar X-ray sources per square degree as a
function of PSPC count rate (\cnts ) in the direction of the Cygnus area. Also 
listed are the relative contributions in percents of various 
spectral types and age groups}

\begin{tabular}{lrrr}
\hline
Limiting PSPC count rate & {3 $10^{-2}$} &  
{1 $10^{-2}$} & {1 $10^{-3}$}  \\ 
\hline 
Nbr of X-ray stars per deg$^{2}$& 0.57 &  2.54 & 45.9  \\
\hline
A     &  3.7 &     3.6 &   3.5      \\
F     & 14.0 &    14.3 &  14.5      \\
G     & 36.2 &    35.5 &  34.6      \\
K     & 25.7 &    25.4 &  25.3     \\
M0-M5 & 20.2 &    21.0 &  21.9     \\
M6-M9 &  0.2 &     0.2 &   0.2     \\
\hline 
 Age $\leq$ 0.15 Gyr               &  62.8   & 61.4  & 59.8   \\
 0.15 Gyr $\leq$ Age $\leq$ 1 Gyr  &  21.7   & 22.4  & 23.2   \\
 Age $\geq$ 1 Gyr                  &  15.5   & 16.2  & 17.0   \\
\hline
\end{tabular}
\label{poptable}
\end{table*}

\begin{table*}

\caption{Observed and predicted number of stellar X-ray sources per square degree and
distribution in spectral types (percentages) for a limiting count rate of 0.02 \cnts .}

\begin{tabular}{lcr}
\hline
                                 & Observed        &  Model \\ 
                                 &                 &prediction \\ \hline
Nbr of X-ray stars per deg$^{2}$ & 1.05 $\pm$ 0.16 & 1.04              \\ \hline
A                                & 15.0 $\pm$ 5.6  &  3.7              \\ 
F                                & 17.5 $\pm$ 6.0  & 14.1              \\
G                                & 25.0 $\pm$ 6.8  & 36.0              \\
K                                & 22.5 $\pm$ 6.6  & 25.5              \\
M0-M5                            & 20.0 $\pm$ 6.3  & 20.5              \\
M6-M9                            &  --             & 0.2               \\
\hline
\end{tabular}
\label{poptable2}
\end{table*}

\subsection{Nature of the unidentified sources}

Early simulations of survey data have shown that for a typical survey exposure the
density of spurious sources was of the order of 0.0145 source per square degree above
ML = 10 (Voges 1995). However, the precise number of false sources in a given survey
field depends on background intensity and structure mainly determined by diffuse
emission in our area. Consequently, the average figure mentioned above may be wrong by
a factor of 2 or more in our case.  The number of spurious sources per square degree
detected above a given ML$_{0}$ may be estimated as:

\begin{displaymath} 
N_{\rm spurious}(\rm ML \geq ML_{0} ) \ = \ 1.45 \times 10^{-2}\ e^{10-\rm ML_{0}}
\end{displaymath}

This relation implies a formal density of spurious sources of 0.0145, 0.11 and 0.29
deg$^{-2}$ for ML$_{0}$ = 10, 8, and 7 respectively. Therefore, only 20\% to 40\% of the 0.53
unidentified sources per square degree with ML $\geq$ 8 may be spurious. The expected
extragalactic contribution computed in section 8 is also too weak to explain the
remaining number of unidentified sources (see Fig. \ref{lognlogs}).

\begin{table*}
\caption{Density of unidentified sources}
\begin{tabular}{rcccccc}
\hline
ML & Flux limit & Total observed& Min-max model& Identified & Unidentified &
Possible range of unidentified \\
   & (\cnts )   & source density& AC density   & non-AC     &              & non-AC non-extragalactic \\
\hline
10 &  0.020     & 1.02          & 0.84 -- 1.41 & 0.10       & 0.10         & 0.0 -- 0.09  \\
8  &  0.012     & 1.74          & 1.32 -- 2.66 & 0.10       & 0.53         & 0.0 -- 0.29  \\
\hline
\end{tabular}
\label{xunid}
\end{table*}

We list in Table \ref{xunid} for two limiting count rates corresponding to the
completeness levels for ML = 10 and ML = 8 several characteristic source densities
(N($\geq$S)\,deg$^{-2}$).  Minimum and maximum model stellar densities were computed
from the error due to binning in age of the XLFs and adding a $\pm$ 1$\sigma$ source
counting error in the 64.5 deg$^{2}$ area. In a similar manner we applied a - 1$\sigma$
counting error to the estimated extragalactic contribution in the test region. The last
column of Table \ref{xunid} lists the maximum possible density range for the optically
unidentified sources which are not associated with active coronae nor with extragalactic
sources. These figures may be further lowered by the unknown fraction of spurious
sources (up to 0.11\,deg$^{-2}$ for ML=8). Using a density of active coronae
extrapolated from the fitted \lnls \ curve instead of the model prediction does not
change the results. Considering the active coronae identified at the 95\% confidence
level on the basis of positional coincidence may allow to identify another 0.14 sources
per square degree above 0.012\,\cnts.  We conclude that there is no evidence for a large
'exotic' population in excess of 0.09\,deg$^{-2}$  and 0.29\,deg$^{-2}$ at 0.02 and
0.012\,\cnts \ respectively and that stars may well account for most if not all of the
unidentified sources.  

\subsection{Constraints on X-ray emission from old neutron stars}

As many as 10$^{9}$ neutron stars could be born in the Galaxy during the last
10$^{10}$ yr.  After a relatively short lived radio emitting episode, these stellar
remnants could reveal themself in the EUV - soft X-ray energy bands if they accrete
from the interstellar medium.

First proposed by Ostriker, Rees \& Silk (1970) the possibility that such a population
could appear in the current X-ray surveys has been recently studied by several
authors. Treves \& Colpi (1991) have discussed the case of magnetized neutron stars
accreting from an homogeneous interstellar medium with particular emphasis put on the
ROSAT PSPC instrument. Blaes \& Madau (1993) have further studied the observability of
old isolated neutron stars in a wide range of energies. They use a slightly different
neutron star velocity distribution as Treves \& Colpi and consider several typical
directions in the Galaxy representing the various distributions of hot tenuous and cool
denser phases observed in the nearby interstellar medium. Colpi et al. 
(1993) have considered in more detail the case of the neutron stars accreting from
dense molecular clouds.

Bondi accretion heavily depends on the relative velocity of the neutron star with
respect to the interstellar medium and only the low velocity tail of the neutron star
velocity distribution is expected to be detectable in X-rays. Another important
parameter is the strength of the magnetic field which channels the accreted matter
towards the poles and increases the effective temperature of the X-ray photons
emerging from the polar cap.

The sensitivity of the ROSAT all-sky survey offers for the first time the possibility
to sample this hypothetical population up to rather large distances over the whole
sky. Treves \& Colpi (1991) find that $\sim$ 5000 old neutron stars spread almost
isotropically over the sky should be present above a PSPC count rate of 0.015\,\cnts \
which is the assumed sensitivity limit of the all-sky survey.  Assuming a total number
of N$_{\rm ns}$ = 10$^{9}$ neutron stars in the Galaxy, Blaes and Madau (1993) argue
that at this level of X-ray flux, the number of detectable X-ray emitting members
could be in the range from $\sim$ 2,000 to 10,000 depending on the accretion mode
(polar or isotropic) with a marked concentration in the galactic plane.

\begin{figure}
\psfig{figure=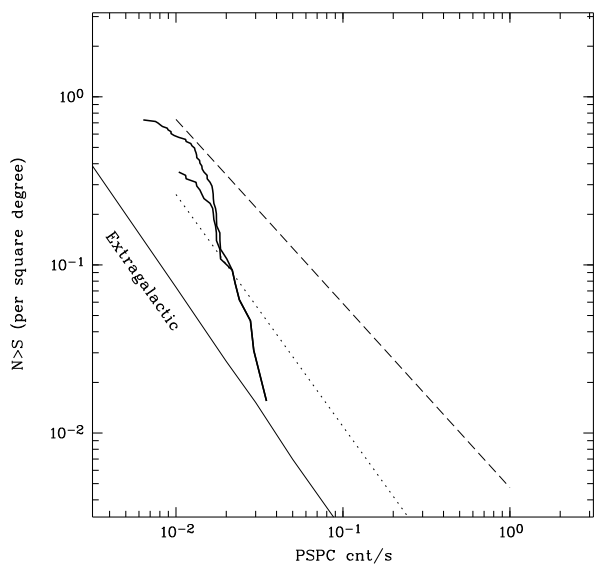,width=8.8cm,bbllx=0cm,bburx=21cm,bblly=2cm
,bbury=20cm}

\caption[]{The \lnls \ relation for all unidentified sources with ML $\geq$ 8 and ML
$\geq$ 10 (thick lines) in the 64.5 square degrees area.  Over-plot are the predictions of
Treves \& Colpi (1991) for an interstellar density of n = 0.07\,cm$^{-3}$ (dotted
line) and n = 1\,cm$^{-3}$ (dashed line)} 

\label{colpi}   
\end{figure}

\begin{figure}
\psfig{figure=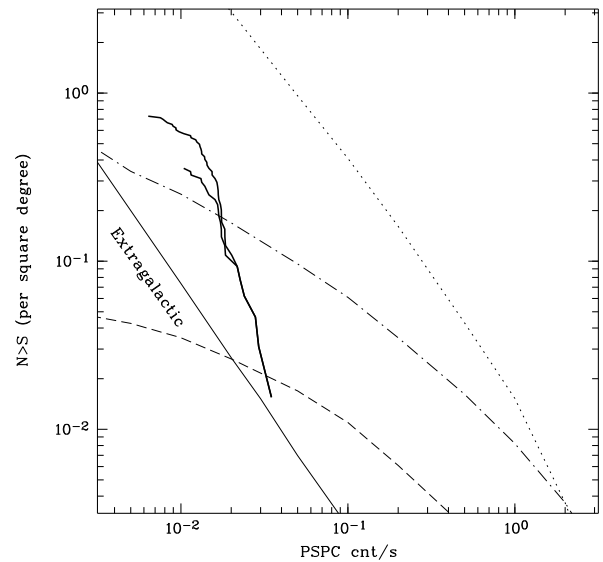,width=8.8cm,bbllx=0cm,bburx=21cm,bblly=2cm,bbury=20cm}
\caption[]{
The \lnls \  relation for all unidentified sources with ML $\geq$ 8 and ML
$\geq$ 10 (thick lines) in the 64.5 square degrees area.
Over-plot are the predictions of Blaes \& Madau (1993); Case 1 polar (dotted line), 
Case 1 isotropic (dotted dashed line) and Case 4 polar (dashed line)}
\label{blaes}   
\end{figure}

We show in Fig. \ref{colpi} and \ref{blaes} the \lnls \ function for our unidentified 
source fraction together with the various theoretical predictions. 

Our observed \lnls \ relation for unidentified sources and the upper limits listed in Table
\ref{xunid} clearly rule out the n = 1\,cm$^{-3}$ density case of Treves \& Colpi and are
marginally compatible with the n = 0.07\,cm$^{-3}$ density case. More stringent constraints
may be put on the model of Blaes \& Madau (1993) (N$_{\rm ns}$ = 10$^{9}$) since we can
exclude their case 1 thought to be representative of the galactic plane whatever is the
accretion mode polar or isotropic.  

\begin{figure}
\psfig{figure=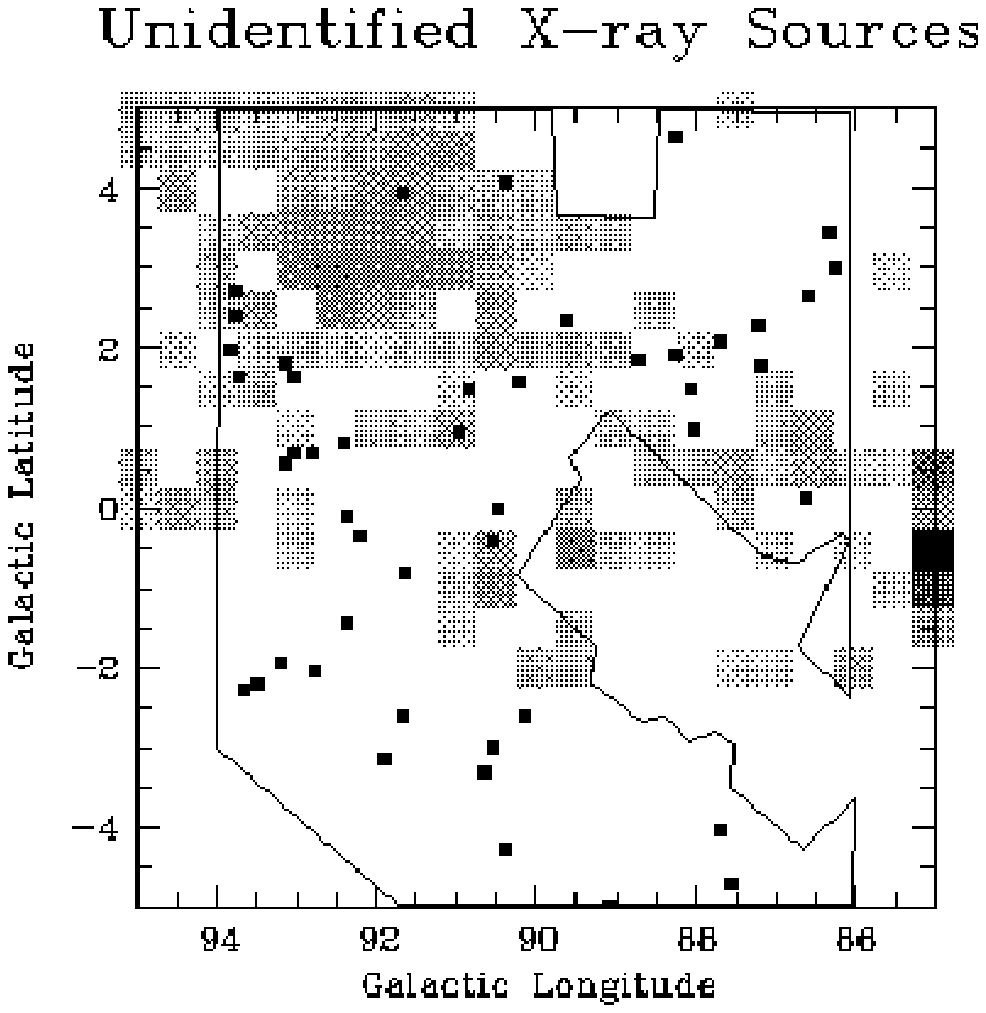,width=8.8cm,bbllx=5cm,bburx=16cm,bblly=8cm,bbury=20cm}

\caption[]{Position of the unidentified X-ray sources (maximum likelihood $\geq$ 8, all
count rates) overlayed on the CO map from Dame et al. (1987) where we also sketch the area
investigated in X-rays. The absence of a correlation between the distribution of the
unidentified population and the CO structures seems to rule out the possibility that these
sources consist of a majority of extragalactic sources or of lonely neutron stars accreting
from the dense interstellar medium. The \lnls \ function and the spatial repartition rather
suggest that most unidentified sources are active coronae}

\label{unidco}   
\end{figure}

The enhanced density of the interstellar medium in molecular clouds favours the detection of
accreting neutron stars in their direction.  Colpi, Campana \& Treves (1993) estimated that
the CYG OB7 cloud which is the dominant CO feature in our area could harbour as many as 1000
lonely accreting neutron stars.  Fig. \ref{comap} shows that we cover a very large fraction
of this molecular cloud, especially the densest parts.  These authors estimate that
$\approx$ 70 neutron stars with accretion luminosity of the order of 10$^{32}$ \ergs \ could
be detected above 0.02\,\cnts. Our detection at similar count rates of X-ray emission from
two bright OB stars probable members of CYG OB7 shows that we are indeed sensitive to
accreting neutron stars located in this molecular cloud. Our observations seem to be at
variance with predictions since at the level of 0.02\,\cnts , we only have 8 unidentified
sources over the whole field (see Table \ref{overallstat1}) instead of the expected 72 for
the entire cloud or 36 for a half cloud.  We show in Fig.  \ref{unidco} the position of the
unidentified sources overlayed on the CO map (Dame et al. 1987). The absence of correlation
between the location of unidentified sources and CO intensity seems indeed incompatible with
the idea that the majority of these sources are old neutron stars accreting from the dense
interstellar medium. However, we cannot completely rule out a conspiracy resulting from the
correlation between high X-ray luminosities and photoelectric absorption. In a recent paper,
Zane et al. (1995) perform a more accurate analysis of the distribution of old neutron star.
Assuming the most favourable cases of polar cap accretion, they predict that $\approx$ 9 to
17 such sources may be detected in the entire CYG OB7 above 0.02\,\cnts. This reduced number
is now compatible with our observations.

Considering the various ingredients entering these models the lack of a large
population of unidentified, old neutron star candidate X-ray sources may have one or
several different interpretations.  

The first important parameter is the density of the interstellar medium in the
direction surveyed here.  Because of the very low density of the local cavity (n$_{H} \
\leq $ 0.015\,cm$^{-3}$) in which the Sun seems presently embedded and which extends at
least up to 50\,pc in most directions (Paresce 1984, Welsh et al. 1994), we do not
expect the presence of nearby X-ray luminous lonely neutron stars (Blaes \& Madau
1993).  At larger distances from the Sun, the mean density of the interstellar medium
reaches more typical values for the Galaxy (n$_{H} \ \approx $ 1.0\,cm$^{-3}$, Blaes \&
Madau 1993). Although there exists some general agreement on the main structure of the
local bubble, namely a large extent toward galactic longitudes $l$ = 200-270\degree ,
the details of the boundary are poorly known.  However, Welsh et al.  (1994) claim that
the bubble radius could be as narrow as 25\,pc toward $l$ = 90\degree \ and Paresce
(1984) find consistent results although with a lower spatial resolution (see also Zane
et al. (1996) for a recent discussion of the local interstellar medium).  The maps of
interstellar extinction from Neckel \& Klare (1980) show that in most directions
concerned here A$_{V}$ smoothly increases to $\approx$ 1 at 1\,kpc and that in the
innermost regions of the plane the interstellar absorption already reaches A$_{V}$
$\approx$ 2.5 at 500\,pc. This indicates a mean hydrogen density in the range of
0.6-3.0\,cm$^{-3}$ up to a few hundreds of parsecs. We conclude that in the direction
surveyed here in X-rays the interstellar medium has a density and a radial distribution
typical of the mid-plane conditions as defined by Blaes \& Madau (1993) and that the
correct models to compare with are those of case 1 (see Fig.  \ref{blaes}).  

Following Madau \& Blaes (1994) we argue that the second crucial parameter is the
velocity distribution of the old neutron stars since the most powerful X-ray sources
will be those few with very low velocities relative to the interstellar medium. If
the mechanism increasing the  velocity dispersion of old stars (Wielen 1977) is also
at work for old neutron stars, the number of low velocity neutron stars may be
severely overestimated.  Finally, the overall number of neutron stars assumed to be
born in the Galaxy may also be overestimated.

Madau \& Blaes (1994) demonstrated that a model taking into account the dynamical
heating of the old neutron star population would indeed drastically decrease the
detectability of these objects in the RASS. Following their computations, our limit on
the density of non-coronal non-extragalactic sources could imply an overall number of
old neutron stars of only (N$_{\rm ns}$ $\approx$ 10$^{8}$) if accretion always occurs
on polar caps. Alternatively, a large number of fossil neutron stars (N$_{\rm ns}$ =
10$^{9}$) would only be acceptable if accretion occurs isotropically and if the
dynamical heating mechanism is at work.  Based on results from the EUVE and WFC
all-sky surveys and using an early report on a parallel optical identification program
of RASS sources carried out by our group in the Taurus constellation (Guillout et al.
1996b), Madau \& Blaes (1994) concluded that the presently available observations may
already hint at a much less numerous old neutron star population than previously
thought. The present constraints, however, do not preclude the discovery of few
relatively close ($\approx$ 100\,pc) isolated neutron stars (e.g. Walter et
al. 1996).

\section{Conclusions}

We present the first results of a systematic identification programme of ROSAT all-sky
survey sources in the galactic plane. The test region analyzed in this paper was chosen to
be typical of deep plane conditions. Centered at $l$ = 90\degree \ $b$ = 0\degree \ it
contains 128 sources above a maximum likelihood of 8 in a 64.5 deg$^{2}$ area.  The overall
\lnls \ function is fully consistent with that derived from the Einstein galactic plane
surveys having similar X-ray sensitivity (Hertz \& Grindlay 1984) and with the results of
deep ROSAT pointings (Morley et al. 1996).  Thanks to a high completeness of optical
identifications we are in a position to discuss the relative contributions of various kinds
of X-ray sources.  An overwhelming fraction of X-ray sources are identified with late type
active coronae.  Our identified sample exhibits the X-ray luminosity distribution expected
in a flux limited survey and we detect F-G stars up to $\approx$ 300\,pc while M stars are
not found beyond $\approx$ 100\,pc. X-ray count models account satisfactorily both for the
observed \lnls \ relation and for the distribution in spectral types and suggest that most
of the stars detected by ROSAT in this low latitude region are probably younger than 1\,Gyr.
We also put strong constraints on the possible contributions of old neutron stars accreting
from the interstellar medium and discuss the possible implications of our results.  

\begin{acknowledgements} We thank A. Strong for kindly providing the CO and HI maps in
electronic form.  We thank the night assistants at Observatoire de Haute-Provence for
carrying out some of the observations at the 1.2m telescope.  The ROSAT project is
supported by the Bundesministerium f\"ur Bildung, Wissenschaft, Forschung und
Technologie (BMBF/DARA) and the Max-Planck-Gesellschaft. C.M. 
acknowledges support from a CNRS-MPG cooperation contract and thanks Prof. J.
Tr\"umper and the ROSAT group for their hospitality and fruitful discussions. This
research has made use of the SIMBAD database operated at CDS, Strasbourg, France.

\end{acknowledgements}

\vskip 2cm

\noindent {\bf Note added in proof:} After the paper was accepted for
publication, a new search in SIMBAD revealed that RX J2134.2+4911 (index 97) is
most probably identified with the dwarf nova V1081 Cyg.

\onecolumn
\begin{table*}

   \caption[]{\label{optid} Optical identifications for sources with ML $\geq$ 8.  Entries
are sorted by decreasing count rates. For the sake of completeness, we also list the
proposed counterparts having a probability of identification in the range 95\% - 98\%.
These more uncertain cases are marked by a '?' as first letter of the identification name
and the class acronym is followed by a '?'}

\end{table*}

\scriptsize
\tablehead{\hline \tbsp
Source              &ROSAT               &Class              &Spectral
   &Identification              &Cnt rate            & Error &Max      &Position 
       &CaII H\&K  \\[0.8ex]
index               &name                &                   &type
   &                            &(cts/s)             &       &lik.     &prob.
       &prob.      \\[0.8ex]
\hline \tbsp}
\tabletail{\hline \multicolumn{10}{l}{\sl Continued on next page}\\ \hline}
\tablelasttail{\hline}
\begin{tabular}{cclllccrcr}
  1&RX J2112.7+5006&WD &DAw   &GD 394              &1.310&0.041&3643.6&1.00&    --\\   
  2&RX J2133.9+4535&AC &G8III &HR 8252             &0.779&0.035&1297.3&1.00&    --\\   
  3&RX J2052.7+4639&WD &DO    &C                   &0.351&0.021& 863.4&1.00&    --\\   
  4&RX J2125.2+4942&AC &K4V   &A =GSC-0359801594   &0.250&0.019& 422.2&1.00&    0.30\\   
  5&RX J2100.8+4530&AC &G3V   &SAO 50350           &0.244&0.017& 425.1&1.00&    0.70\\ [0.8ex]  
  6&RX J2106.0+5421&AC &F8V   &HD 235440           &0.140&0.017& 106.2&1.00&$<$ 0.04\\   
  7&RX J2104.1+4912&AC &M4e   &B                   &0.139&0.014& 267.3&1.00&    --\\   
  8&RX J2130.8+4827&AC &G3V   &SAO 50961           &0.136&0.015& 168.7&1.00&$<$ 0.18\\   
  9&RX J2120.9+4636&AC &F9V   &A =GSC-0358903858   &0.114&0.013& 150.5&1.00&    0.64\\   
 10&RX J2124.7+4639&AC &K3+K7V&A+B                 &0.113&0.013& 159.4&1.00&    0.59\\ [0.8ex]  
 11&RX J2117.3+5044&WD &DA    &B =LAN 121          &0.109&0.013& 105.3&1.00&    --\\   
 12&RX J2135.9+4728&AGN&Seyf 1&B                   &0.101&0.012& 160.4&1.00&    --\\   
 13&RX J2102.6+4552&AC &K3V   &HD 200560           &0.095&0.011& 168.8&1.00&    --\\   
 14&RX J2123.1+4831&AC &F0V   &HD 203839           &0.077&0.011&  81.2&1.00&$<$ 0.92\\   
 15&RX J2055.3+5025&AC &G0V   &A =GSC-0358301038   &0.070&0.012&  58.8&1.00&$<$ 0.01\\ [0.8ex]  
 16&RX J2109.2+4810&AC &K7V   &A =GSC-0359205781   &0.070&0.011&  68.3&0.98&    0.85\\   
 17&RX J2100.9+5103&AC &M2Ve  &A =G 231 -24        &0.062&0.011&  51.8&1.00&    --\\   
 18&RX J2049.6+5119&AC &F2V   &HD 198638           &0.061&0.011&  46.4&1.00&    --\\   
 19&RX J2059.3+5303&AC &M3.5Ve&A =GSC-0395201062   &0.061&0.012&  38.8&1.00&    --\\   
 20&RX J2056.7+4940&AC &A7V   &SAO 50269           &0.060&0.010&  61.7&1.00&    --\\   [0.8ex]
 21&RX J2123.1+5021&AC &G0V   &A =BD+49 3512       &0.059&0.010&  52.2&1.00&    0.19\\   
 22&RX J2107.8+4932&AC &G8V   &A =GSC-0359600261   &0.049&0.009&  44.6&1.00&    0.77\\   
 23&RX J2100.1+4841&AC &G9V   &BD+48 3260 E        &0.048&0.009&  37.4&0.99&    0.44\\   
 24&RX J2104.7+5223&AC &G2V   &A  =GSC-0360000513  &0.048&0.011&  33.0&0.99&    0.04\\   
 25&RX J2044.6+4758&AC &      &A =GSC-0357800872   &0.048&0.009&  49.6&0.99&    --\\   [0.8ex]
 26&RX J2107.3+5202&AC &F8V   &V1061 Cyg =HD 235444&0.047&0.010&  40.1&1.00&    --\\   
 27&RX J2052.3+4820&AC &G9V   &A =GSC-0357901469   &0.045&0.009&  43.8&0.99&    0.64\\   
 28&RX J2118.4+4356&OB &O8Ve  &HR 8154             &0.045&0.009&  48.6&1.00&    --\\   
 29&RX J2109.3+5138&AC &G3V   &A =GSC-0360000175   &0.041&0.009&  29.7&1.00&    0.53\\   
 30&RX J2120.4+4733&AC &K4V   &HD 203418           &0.040&0.008&  34.9&1.00&    0.91\\ [0.8ex]  
 31&RX J2104.2+5015&AC &K7e   &A                   &0.038&0.008&  16.0&1.00&    --\\   
 32&RX J2057.3+4813&AC &A2m   &A =GSC-0357900338   &0.037&0.008&  33.6&0.99&$<$ 0.11\\   
 33&RX J2119.0+5207&AC &G1V   &A =GSC-0360101235   &0.036&0.009&  32.3&0.99&    0.20\\   
 34&RX J2117.8+5112&AC &A2V   &HD 203028           &0.035&0.008&  20.7&1.00&    --\\   
 35&RX J2133.3+4726&?? &      &                    &0.035&0.008&  41.8&0.00&    --\\   [0.8ex]
 36&RX J2134.0+4525&AC &G3V   &A =GSC-0359101945   &0.034&0.010&  16.0&0.99&    0.60\\   
 37&RX J2103.4+5021&AC &K0III &HR 8072             &0.034&0.008&  35.0&1.00&    0.96\\   
 38&RX J2040.6+4859&AC &F9V   &SAO 49928           &0.033&0.008&  38.4&1.00&$<$ 0.17\\   
 39&RX J2102.9+4854&AC &G3V   &A =GSC-0359600374   &0.033&0.008&  24.1&0.99&    0.49\\   
 40&RX J2113.3+5140&AC &M4Ve  &A                   &0.032&0.008&  42.9&1.00&    --\\   [0.8ex]
 41&RX J2123.5+4621&AC &K4V   &A  =GSC-0359002325  &0.032&0.008&  28.1&0.98&    0.63\\   
 42&RX J2100.9+4857&AC &K2V   &A =GSC-0359601318   &0.032&0.008&  23.5&0.00&    0.08\\   
 43&RX J2130.1+4901&AC &M5Ve  &B                   &0.032&0.008&  26.6&1.00&    0.03\\   
 44&RX J2054.1+4942&AC &F8V   &A  =GSC-0358300309  &0.032&0.009&  14.8&0.99&$<$ 0.01\\   
 45&RX J2116.6+4645&AC &M5Ve  &A                   &0.032&0.008&  29.7&1.00&    0.28\\   [0.8ex]
 46&RX J2110.2+5333&OB &Ap    &HR 8106             &0.031&0.010&  18.4&1.00&    --\\   
 47&RX J2125.3+4642&AC &F0V   &HR 8208             &0.031&0.008&  26.0&1.00&    --\\   
 48&RX J2118.5+5247&AC &      &A =GSC-0395301190   &0.030&0.008&  23.0&0.99&    --\\   
 49&RX J2128.7+4409&?? &      &                    &0.029&0.009&  10.4&0.00&    --\\   
 50&RX J2121.7+5049&AC &A0V   &HD 235498           &0.029&0.008&  28.0&1.00&$<$ 0.95\\ [0.8ex]  
 51&RX J2132.5+4849&?? &      &                    &0.028&0.008&  19.7&0.94&$<$ 0.02\\   
 52&RX J2122.4+5023&AC &G9V   &A  =GSC-0359700228  &0.028&0.008&  22.2&0.97&    0.46\\   
 53&RX J2035.9+4900&AC &G5-8V &A =GSC-0358101856   &0.027&0.007&  22.5&0.99&    0.75\\   
 54&RX J2108.2+5313&AC &A0V   &HD 201543           &0.027&0.009&  15.2&1.00&    --\\   
 55&RX J2128.6+4653&AC &K5V   &A =GSC-0359400075   &0.026&0.007&  21.7&0.93&    0.59\\ [0.8ex]  
 56&RX J2048.0+4903&AC &      &A =GSC-0358200586   &0.026&0.007&  27.5&0.99&    --\\   
 57&RX J2115.4+4437&AC &K0V   &A =GSC-0318101403   &0.025&0.007&  20.7&0.98&    0.98\\   
 58&RX J2109.9+4809&AC &A5V   &SAO 50523           &0.025&0.007&  16.1&1.00&    --\\   
 59&RX J2108.6+4927&AC &M2Ve  &A                   &0.024&0.007&  22.9&1.00&    0.17\\   
 60&RX J2050.8+4743&?? &      &                    &0.024&0.007&  21.0&0.84&    --\\   [0.8ex]
 61&RX J2135.6+4523&AC &F9V   &A =GSC-0359102803   &0.023&0.008&  10.6&0.00&    0.06\\   
 62&RX J2119.9+5227&AC?&      &? A =GSC-0360101367 &0.023&0.007&  23.4&0.98&    --\\   
 63&RX J2100.3+5219&AC &M1Ve  &A =GSC-0360000514   &0.022&0.008&  19.0&1.00&    --\\   
 64&RX J2056.3+4817&?? &      &                    &0.022&0.007&  13.5&0.00&    --\\   
 65&RX J2127.3+5121&AC &      &A+B                 &0.020&0.007&  21.1&0.99&    --\\   [0.8ex]
 66&RX J2041.9+4746&?? &      &                    &0.020&0.007&   8.4&0.00&    --\\   
 67&RX J2126.3+4654&AC &      &A =GSC-0359401755   &0.020&0.007&  10.2&0.98&    --\\   
 68&RX J2055.8+5044&AC &G8V   &A =GSC-0358700365   &0.019&0.007&  15.1&0.97&    0.23\\   
 69&RX J2117.5+5127&AC &M0Ve  &A =GSC-0360100450   &0.019&0.007&  12.8&1.00&    --\\   
 70&RX J2057.7+4752&?? &      &                    &0.018&0.006&  11.3&0.00&    --\\   [0.8ex]
 71&RX J2131.2+4533&?? &      &                    &0.018&0.007&   9.3&0.00&    --\\   
 72&RX J2054.6+5120&?? &      &                    &0.018&0.006&  12.0&0.93&$<$ 0.01\\   
 73&RX J2121.5+4317&?? &      &                    &0.018&0.006&   8.2&0.00&    --\\   
 74&RX J2128.4+4900&?? &      &                    &0.017&0.006&  10.5&0.00&    --\\   
 75&RX J2058.5+4836&?? &      &                    &0.017&0.006&  10.7&0.92&    --\\   [0.8ex]
 76&RX J2130.0+4740&AC?&K4V   &? A =Plat 2137      &0.017&0.006&  12.6&0.97&    --\\   
 77&RX J2106.2+4437&AC &M5Ve  &A                   &0.017&0.006&  16.7&1.00&    0.33\\   
 78&RX J2128.5+4626&?? &      &                    &0.017&0.007&  11.1&0.81&    --\\   
 79&RX J2117.7+5139&?? &      &                    &0.017&0.006&  23.3&0.00&    --\\   
 80&RX J2101.6+4730&AC &G0V   &HD 200406           &0.017&0.006&  10.1&1.00&$<$ 1.00\\ [0.8ex]  
 81&RX J2046.7+4728&AC?&      &? A =GSC-0357800892 &0.017&0.006&   9.2&0.98&    --\\   
 82&RX J2136.6+4911&?? &      &                    &0.017&0.006&  19.0&0.00&    --\\   
 83&RX J2118.6+5010&AC &K0V   &HD 203136           &0.017&0.005&  19.1&1.00&    --\\   
 84&RX J2130.3+4709&CV &Me+DA &B                   &0.016&0.006&  10.6&1.00&    --\\   
 85&RX J2116.0+4827&?? &      &                    &0.016&0.006&   8.9&0.00&    --\\   [0.8ex]
 86&RX J2124.7+4714&AC &      &A =GSC-0359402027   &0.016&0.006&  12.2&0.99&    --\\   
 87&RX J2117.5+4330&AC &F8V   &A =GSC-0318100914   &0.016&0.006&   8.7&0.99&    0.73\\   
 88&RX J2137.6+4916&?? &      &                    &0.016&0.005&  15.3&0.90&    --\\   
 89&RX J2100.6+5039&AC &      &A =GSC-0360000197   &0.016&0.006&   9.1&0.99&    --\\   
 90&RX J2110.5+4913&AC &      &A =GSC-0359601415   &0.016&0.006&   8.6&0.99&    --\\   [0.8ex]
 91&RX J2121.5+4732&AC &M0Ve  &A =GSC-0359303030   &0.016&0.006&   8.4&1.00&    0.95\\   
 92&RX J2052.8+4723&?? &      &                    &0.015&0.006&   8.0&0.00&    --\\   
 93&RX J2118.6+5039&?? &      &                    &0.015&0.006&   8.3&0.00&    --\\   
 94&RX J2122.3+4730&AC &M4.5Ve&A =GSC-0359305051   &0.015&0.006&  10.4&1.00&    --\\   
 95&RX J2050.4+4913&AC &      &A =GSC-0358300408   &0.015&0.006&   9.3&0.99&    --\\   [0.8ex]
 96&RX J2113.0+4834&AC &G1V   &A =GSC-0359303505   &0.015&0.006&   9.9&0.98&    0.83\\   
 97&RX J2134.2+4911&AC?&M2-M6 &? A =Plat 5991      &0.015&0.005&  21.7&0.95&    --\\   
 98&RX J2119.5+4351&?? &      &                    &0.014&0.006&   8.9&0.72&    --\\   
 99&RX J2117.4+5152&?? &      &                    &0.014&0.005&  13.1&0.00&    --\\   
100&RX J2123.2+5057&?? &      &                    &0.014&0.005&  10.3&0.00&    --\\   [0.8ex]
101&RX J2053.4+4759&?? &      &                    &0.014&0.006&   9.0&0.00&    --\\   
102&RX J2124.3+5059&AC &K7Ve  &A                   &0.013&0.005&  17.3&1.00&    --\\   
103&RX J2114.1+4840&AC?&      &? A+B               &0.013&0.005&  14.4&0.97&    --\\   
104&RX J2043.7+4727&AC?&      &? A =GSC-0357801789 &0.013&0.006&   8.8&0.97&    --\\   
105&RX J2122.6+4956&?? &      &                    &0.013&0.005&  11.7&0.00&    --\\   [0.8ex]
106&RX J2103.6+4845&AC &      &A =GSC-0359600092   &0.013&0.005&   9.7&0.99&    --\\   
107&RX J2059.8+4937&AC?&      &? A =GSC-0358300546 &0.013&0.005&   8.8&0.96&    --\\   
108&RX J2126.9+4636&?? &      &                    &0.013&0.005&   8.1&0.00&    --\\   
109&RX J2120.2+5127&AC &      &A =GSC-0360100440   &0.013&0.006&  12.1&1.00&    --\\   
110&RX J2130.7+4919&AC &F9    &A =Plat 2930        &0.012&0.005&  11.7&0.94&    0.08\\ [0.8ex]  
111&RX J2123.7+4636&?? &      &                    &0.012&0.005&   8.7&0.94&    --\\   
112&RX J2043.4+5002&AC?&      &? A =GSC-0358200283 &0.012&0.005&   8.2&0.96&    --\\   
113&RX J2107.6+5048&AC &G1V   &A  =GSC-0360000137  &0.012&0.005&  12.9&0.99&$<$ 0.11\\   
114&RX J2117.5+5243&AC?&      &? A =GSC-0395301075 &0.012&0.004&  18.3&0.97&    --\\   
115&RX J2120.8+5209&?? &      &                    &0.011&0.005&  12.3&0.94&    --\\   [0.8ex]
116&RX J2121.8+5135&AC &      &A =GSC-0360100693   &0.010&0.005&   9.2&0.99&    --\\   
117&RX J2121.0+5049&?? &      &                    &0.010&0.004&  10.4&0.92&    --\\   
118&RX J2058.1+4552&AC?&      &? A =GSC-0357501843 &0.010&0.005&   9.4&0.98&    --\\   
119&RX J2112.9+5314&AC &      &A =GSC-0395300355   &0.010&0.005&   8.8&1.00&    --\\   
120&RX J2122.6+4856&AC?&      &? A =GSC-0359800927 &0.009&0.004&   8.1&0.98&    --\\   [0.8ex]
121&RX J2122.0+5059&?? &      &                    &0.009&0.004&   9.0&0.00&    --\\   
122&RX J2105.7+4933&?? &      &                    &0.009&0.004&   8.0&0.00&    --\\   
123&RX J2116.0+5254&?? &      &                    &0.009&0.004&   8.3&0.00&    --\\   
124&RX J2059.6+4731&AC?&      &? A =GSC-0357901518 &0.008&0.004&   8.4&0.95&    --\\   
125&RX J2100.8+5213&AC?&      &? A =GSC-0360000242 &0.008&0.004&   8.9&0.97&    --\\   [0.8ex]
126&RX J2111.8+4942&?? &      &                    &0.008&0.004&   9.2&0.00&    --\\   
127&RX J2108.9+4958&?? &      &                    &0.008&0.004&   8.1&0.00&    --\\   
128&RX J2122.9+4941&?? &      &                    &0.006&0.003&   9.8&0.00&    --\\   
\end{tabular}

\newpage

\onecolumn
\begin{table*}
   \caption[]{\label{optids} Optical identifications for sources with 7 $\leq$ ML $\leq$ 8.
Entries are sorted by decreasing count rates. For the sake of completeness, we also list the
proposed counterparts having a probability of identification in the range 95\% - 98\%.
These more uncertain cases are marked by a '?' as first letter of the identification name
and the class acronym is followed by a '?'}
\end{table*}

\scriptsize
\tablehead{\hline \tbsp
Source              &ROSAT               &Class              &Spectral
   &Identification              &Cnt rate            & Error &Max      &Position 
       &CaII H\&K  \\[0.8ex]
index               &name                &                   &type
   &                            &(cts/s)             &       &lik.     &prob.
       &prob.      \\[0.8ex]
\hline \tbsp}
\tabletail{\hline \multicolumn{10}{l}{\sl Continued on next page}\\ \hline}
\tablelasttail{\hline}
\begin{tabular}{cclllccrcr}
129&RX J2053.2+4951&?? &      &                    &0.020&0.008&   7.9&0.00&    --\\   
130&RX J2054.0+4858&?? &      &                    &0.019&0.007&   7.7&0.00&    --\\   
131&RX J2049.6+5128&AC &      &A =GSC-0358700147   &0.017&0.008&   7.3&0.99&    --\\   
132&RX J2052.3+4843&AC &      &A =GSC-0357901214   &0.016&0.006&   7.1&0.99&    --\\   
133&RX J2129.2+4604&?? &      &                    &0.016&0.006&   7.9&0.00&    --\\  [0.8ex] 
134&RX J2124.5+5133&?? &      &                    &0.015&0.006&   7.0&0.00&    --\\   
135&RX J2117.2+5241&AC &      &A =GSC-0395301223   &0.014&0.006&   7.9&0.99&    --\\   
136&RX J2057.6+4750&?? &      &                    &0.013&0.006&   7.8&0.00&    --\\   
137&RX J2111.6+4809&CV &      &V1500 Cyg           &0.013&0.005&   7.6&1.00&    --\\   
138&RX J2051.8+4741&?? &      &                    &0.013&0.005&   7.7&0.00&    --\\   [0.8ex]
139&RX J2057.1+4929&?? &      &                    &0.013&0.005&   7.9&0.00&    --\\   
140&RX J2043.1+4752&AC?&      &? A =GSC-0357802133 &0.013&0.005&   7.4&0.98&    --\\   
141&RX J2127.1+4801&?? &      &                    &0.012&0.005&   7.5&0.00&    --\\   
142&RX J2048.8+4631&?? &      &                    &0.012&0.005&   7.8&0.94&    --\\   
143&RX J2110.5+5244&?? &      &                    &0.011&0.005&   7.8&0.00&    --\\   [0.8ex]
144&RX J2052.3+4656&?? &      &                    &0.011&0.005&   7.5&0.30&    --\\   
145&RX J2101.2+4609&OB &B1Ve  &HD 200310           &0.011&0.005&   7.7&1.00&    --\\   
146&RX J2055.7+4815&?? &      &                    &0.011&0.005&   7.3&0.00&    --\\   
147&RX J2045.1+4805&AC &      &A =GSC-0357800263   &0.011&0.005&   8.0&0.99&    --\\   
148&RX J2110.3+4832&?? &      &                    &0.011&0.004&   7.5&0.00&    --\\   [0.8ex]
149&RX J2131.2+4812&?? &      &                    &0.010&0.005&   7.5&0.94&    --\\   
150&RX J2058.7+4845&?? &      &                    &0.010&0.004&   8.0&0.00&    --\\   
151&RX J2100.3+4810&?? &      &                    &0.010&0.004&   7.9&0.89&    --\\   
152&RX J2130.8+4842&?? &      &                    &0.009&0.004&   7.1&0.00&    --\\   
153&RX J2052.6+4703&?? &      &                    &0.009&0.004&   7.5&0.00&    --\\   [0.8ex]
154&RX J2055.0+5039&?? &      &                    &0.009&0.004&   7.4&0.00&    --\\   
155&RX J2121.1+4907&?? &      &                    &0.008&0.004&   7.2&0.95&    --\\   
156&RX J2122.4+4709&?? &      &                    &0.008&0.004&   7.3&0.93&    --\\   
157&RX J2054.6+4903&?? &      &                    &0.008&0.004&   7.8&0.00&    --\\   
158&RX J2121.5+4951&?? &      &                    &0.006&0.003&   7.5&0.00&    --\\   
\end{tabular}

\end{document}